%% file: HIG-15-010_temp.tex
\pdfoutput=1

\documentclass[11pt,twoside,a4paper,cmspaper,final,collab]{cms-tdr}
\def\svnVersion{397251}\def\cmsCernNoTag{CERN-EP/2016-125}\def\cmsCernDate{\today}\def\cmsMessage{Published in the Journal of High Energy Physics as \href{http://dx.doi.org/10.1007/JHEP03(2017)032}{\doi{10.1007/JHEP03(2017)032.}}}

\begin{document}\cmsNoteHeader{HIG-15-010}

\hyphenation{had-ron-i-za-tion}
\hyphenation{cal-or-i-me-ter}
\hyphenation{de-vices}
\RCS$Revision: 378582 $
\RCS$HeadURL: svn+ssh://svn.cern.ch/reps/tdr2/papers/HIG-15-010/trunk/HIG-15-010.tex $
\RCS$Id: HIG-15-010.tex 378582 2016-12-21 13:01:44Z lviliani $
\newlength\cmsFigWidth
\ifthenelse{\boolean{cms@external}}{\setlength\cmsFigWidth{0.85\columnwidth}}{\setlength\cmsFigWidth{0.4\textwidth}}
\ifthenelse{\boolean{cms@external}}{\providecommand{\cmsLeft}{top\xspace}}{\providecommand{\cmsLeft}{left\xspace}}
\ifthenelse{\boolean{cms@external}}{\providecommand{\cmsRight}{bottom\xspace}}{\providecommand{\cmsRight}{right\xspace}}

\cmsNoteHeader{HIG-15-010}

\newcommand{\WW}{\ensuremath{\PWp\PWm}\xspace}
\newcommand{\hww}{\ensuremath{\PH\to\WW}\xspace}
\newcommand{\hwwllvv}{\ensuremath{\PH \to\WW \to 2\ell 2\nu}\xspace}
\newcommand{\wwllvv}{\ensuremath{\WW \to 2\ell2\nu}\xspace}

\newcommand{\mt}      {\ensuremath{m_\mathrm{T}}\xspace}

\newcommand{\dytt}{\ensuremath{\Z/\gamma^*\to\tau^+\tau^-}\xspace}
\newcommand{\dymm}{\ensuremath{\Z/\gamma^*\to\mu^+\mu^-}\xspace}

\newcommand{\mll}{\ensuremath{m_{\ell\ell}}\xspace}
\newcommand{\pth}{\ensuremath{p_{\mathrm{T}}^{\mathrm{H}}}\xspace}

\newcommand{\x}{\ensuremath{\phantom{0}}}
\newcommand{\xx}{\ensuremath{\phantom{00}}}

\title{Measurement of the transverse momentum spectrum of the Higgs boson produced in pp collisions at $\sqrt{s} = 8\TeV$ using $\PH\to\PW\PW$ decays}

\date{\today}

\abstract{
The cross section for Higgs boson production in pp collisions is studied using the $\PH \to \PW^+ \PW^-$  decay mode, followed by leptonic decays of the W bosons to an oppositely charged electron-muon pair in the final state. The measurements are performed using data collected by the CMS experiment at the LHC at a centre-of-mass energy of 8\TeV, corresponding to an integrated luminosity of 19.4\fbinv. The Higgs boson transverse momentum ($\pt$) is reconstructed using the lepton pair \pt and missing \pt. The differential cross section times branching fraction is measured as a function of the Higgs boson \pt in a fiducial phase space defined to match the experimental acceptance in terms of the lepton kinematics and event topology. The production cross section times branching fraction in the fiducial phase space is measured to be $39\pm 8\stat \pm 9\syst\unit{fb}$. The measurements are found to agree, within experimental uncertainties, with theoretical calculations based on the standard model.
}

\hypersetup{%
pdfauthor={CMS Collaboration},%
pdftitle={Measurement of the transverse momentum spectrum of the Higgs boson produced in pp collisions at sqrt(s) = 8 TeV using the H to WW decays},%
pdfsubject={CMS},%
pdfkeywords={CMS, physics, Higgs boson}}

\maketitle

\section{Introduction}
\label{sec:introduction}
The discovery of a new boson at the CERN LHC reported by the ATLAS and CMS collaborations~\cite{Aad:2012tfa,Chatrchyan:2012ufa,Chatrchyan:2013lba} has been followed by a comprehensive
set of measurements aimed at establishing the properties of the new boson.
Results reported by ATLAS and CMS~\cite{Aad:2013wqa, Aad:2013xqa, Chatrchyan:2013zna, Chatrchyan:2013iaa, Chatrchyan:2013mxa,Chatrchyan:2014nva, Chatrchyan:2014vua, Khachatryan:2014ira, Aad:2014eha, Aad:2014eva, Aad:2014xzb, Khachatryan:2014kca, ATLAS:2014aga, Khachatryan:2014jba, Aad:2015vsa, Aad:2015rwa, Aad:2015zhl, Aad:2015gba,Khachatryan:2016ctc}, so far, are consistent with the standard model (SM) expectations for the Higgs boson ($\PH$).

Measurements of the production cross section of the Higgs boson times branching fraction in a restricted part of the phase space (fiducial phase space) and its kinematic properties represent an important test for possible deviations from the SM predictions.
In particular, it has been shown that the Higgs boson transverse momentum
($\pth$) spectrum can be significantly affected by the presence of interactions not predicted by the SM~\cite{Harlander:2013oja, Banfi:2013yoa,Azatov:2013xha, Grojean:2013nya, Dawson:2014ora}.
In addition, these measurements allow accurate tests of the theoretical calculations in the SM Higgs sector, which offer up to next-to-next-to-leading-order (NNLO) accuracy in perturbative Quantum ChromoDynamics (pQCD), up to next-to-next-to-leading-logarithmic (NNLL) accuracy in the resummation of soft-gluon effects at small \pt, and up to next-to-leading-order (NLO) accuracy in perturbative electroweak corrections~\cite{Bozzi:2005wk,deFlorian:2012mx,Grazzini:2013mca}.

Measurements of the fiducial cross sections and of several differential
distributions, using the $\sqrt{s}=8\TeV$ LHC data, have been reported by ATLAS~\cite{Aad:2014tca,Aad:2014lwa,Aad:2015lha} and CMS~\cite{Khachatryan:2015rxa,Khachatryan:2015yvw} for the ${\PH \to \Z\Z \to 4\ell}$ ($\ell = \Pe,\mu$) and $\PH \to \gamma\gamma$ decay channels, and recently by ATLAS~\cite{Aad:2016lvc} for the ${\PH\to{}\WW\to \Pe^{\pm} \mu^{\mp}\nu\nu}$ decay channel. In this paper we report a measurement of the fiducial cross section times branching fraction ($\sigma \times \mathcal{B}$) and \pt{} spectrum for Higgs boson production in ${\PH\to{}\WW\to \Pe^{\pm} \mu^{\mp}\nu\nu}$~decays, based on $\sqrt{s} = 8\TeV$ LHC data.
The analysis is performed looking at different flavour leptons in the final state in order to suppress the sizeable contribution of backgrounds containing a same-flavour lepton pair originating from Z boson decay.
Although the \hwwllvv{} channel has lower resolution in the \pth{} measurement
compared to the $\PH \to \gamma\gamma$ and  $\PH \to \Z\Z\to 4\ell$ channels
because of neutrinos in the final state, the channel has a significantly
larger $\sigma \, \mathcal{B}$, exceeding those for $\PH \to \gamma\gamma$ by a factor
of 10 and $\PH \to \Z\Z\to 4\ell$ by a factor of 85 for a Higgs boson mass of
125\GeV~\cite{Heinemeyer:2013tqa}, and is characterized by good signal
sensitivity. Such sensitivity allowed the observation of a Higgs boson at the level of 4.3 (5.8 expected)
standard deviations for a mass hypothesis of 125.6\GeV using the full LHC data set at 7 and 8\TeV~\cite{Chatrchyan:2013iaa}.

The measurement is performed in a fiducial phase space defined by kinematic requirements on
the leptons that closely match the experimental event selection.
The effect of the limited detector resolution, as well as the
selection efficiency with respect to the fiducial phase space are corrected to
particle level with an unfolding procedure~\cite{Cowan:2002in}. This procedure
is based on the knowledge of the detector response matrix, derived from the simulation of the CMS
response to signal events, and consists of an inversion of the response matrix
with a regularization prescription to tame unphysical statistical fluctuations in the
unfolded result.

The analysis presented here is based on the previously published \hwwllvv{}
measurements by CMS~\cite{Chatrchyan:2013iaa}. A notable difference from those
measurements is that this analysis is inclusive in the number of jets, which allows the uncertainties related to the theoretical modelling of additional jets produced in association with the Higgs boson to be reduced. There are two important backgrounds: for \pth{} values below approximately 50\GeV the dominant background is
$\PW\PW$ production, while above 50\GeV the production of top--anti-top
($\ttbar$) quarks dominates.

This paper is organized as follows: In Section~\ref{sec:cms} a brief description of the CMS detector is given. The data sets and Monte Carlo (MC) simulated samples are described in Section~\ref{sec:samples}. The strategy adopted in the analysis is described in Section~\ref{sec:AnalysisStrategy}, including the definition of the fiducial phase space. The event selection and a description of all relevant backgrounds are given in Section~\ref{sec:selections}, followed by an overview of the systematic uncertainties important for the analysis in Section~\ref{sec:systematicUncertainties}.
The technique used for the extraction of the Higgs boson signal contribution is described in Section~\ref{sec:signalExtraction}, together with the signal and background yields and the reconstructed \pth{} spectrum.
The unfolding procedure used to extrapolate the reconstructed spectrum to the fiducial phase space is described in Section~\ref{sec:unfoldingAndSyst}, including a detailed description of the treatment of systematic uncertainties in the unfolding. Finally, Section~\ref{sec:results} presents the result of the measurement of the fiducial $\sigma \, \mathcal{B}$ and \pth{} spectrum, and their comparison with the theoretical predictions.

\section{The CMS experiment}
\label{sec:cms}

The central feature of the CMS apparatus is a superconducting solenoid of 6\unit{m} internal diameter providing a magnetic field of 3.8 T. Within the solenoid volume are a silicon pixel and strip tracker, which cover a pseudorapidity ($\eta$) region of $\abs{\eta}<2.5$, a lead tungstate crystal electromagnetic calorimeter (ECAL), and a brass and scintillator hadron calorimeter (HCAL), each composed of a barrel and two endcap sections, covering $\abs{\eta}<3$. Forward calorimetry extends the $\eta$ coverage provided by the barrel and endcap detectors from $\eta > 3$ to $\eta < 5.2$. Muons are measured in gas-ionization detectors embedded in the steel flux-return yoke outside the solenoid.
A more detailed description of the CMS detector, together with a definition of the coordinate system used and the relevant kinematic variables, can be found in Ref.~\cite{Chatrchyan:2008zzk}.

The particle-flow event algorithm reconstructs and identifies each individual
particle with an optimized combination of information from the various
elements of the CMS
detector~\cite{CMS-PAS-PFT-09-001,CMS-PAS-PFT-10-001,CMS-PAS-PFT-10-002,CMS-PAS-PFT-10-003,Chatrchyan:2011ds}.
The energy of photons is obtained from the ECAL measurement, corrected for
instrumental effects. The energy of electrons is determined from a combination of the electron momentum at the primary interaction vertex as determined by the tracker, the energy of the corresponding ECAL cluster, and the energy sum of all bremsstrahlung photons spatially compatible with originating from the electron track~\cite{Khachatryan:2015hwa}. The momentum of muons is obtained from the curvature of the corresponding track. The energy of charged hadrons is determined from a combination of their momentum measured in the tracker and the matching ECAL and HCAL energy deposits, corrected for zero-suppression effects and for the response function of the calorimeters to hadronic showers. Finally, the energy of neutral hadrons is obtained from the corresponding corrected ECAL and HCAL energy.
Jets are reconstructed from the individual particles using the anti-$k_t$ clustering algorithm with a distance parameter of 0.5, as implemented in the \textsc{fastjet} package~\cite{Cacciari:2008gp, Cacciari:2011ma}.

The missing transverse momentum vector \ptvecmiss is defined as the projection of the negative vector sum of the momenta of all reconstructed particles in an event on the plane perpendicular to the beams. Its magnitude is referred to as the missing transverse energy \MET.

Details on the experimental techniques for the reconstruction, identification,
and isolation of electrons, muons and jets, as well as on the efficiencies of
these techniques can be found in
Refs.~\cite{Khachatryan:2015hwa,CMS:2011aa,Chatrchyan:2012xi,Chatrchyan:2013sba,Khachatryan:2015iwa,Khachatryan:2014gga,Chatrchyan:2011ds}. Details on the procedure used to calibrate the leptons and jets in this analysis can be found in Ref.~\cite{Chatrchyan:2013iaa}.

\section{Data and simulated samples}
\label{sec:samples}

This analysis makes use of the same data and MC simulated samples as those used in the previous $\hww$ study~\cite{Chatrchyan:2013iaa}. Data were recorded by the CMS experiment during 2012 and correspond to an integrated luminosity of 19.4\fbinv at a centre-of-mass energy of 8\TeV. The events are triggered by requiring the presence of either one or a combination of electron
and muon with high \pt and tight identification and isolation criteria. Single-lepton triggers are characterized by \pt thresholds varying from 17 to 27\GeV for electrons and from 17 to 24\GeV for muons. Dilepton $\Pe\mu$ triggers are required to have one electron or one muon with $\pt>17\GeV$ and the other muon or electron with $\pt>8\GeV$. The average combined trigger efficiency for signal events that pass the full event selection is measured to be about $96\%$ in the $\Pe\mu$ final state for a Higgs boson mass of 125\GeV.

The signal and background processes relevant for this analysis are simulated using several MC programs. Simulations of the Higgs boson production through the gluon fusion (ggH) and vector boson fusion (VBF) mechanisms are performed using the first version of the \POWHEG generator (\POWHEG V1)~\cite{Kramer:2005hw,Frixione:2007vw,Lavesson:2008ah,Alioli:2008tz, Nason:2009ai} with NLO accuracy in pQCD, while \textsc{Pythia 6.426}~\cite{Sjostrand:2006za} is used to simulate associated Higgs boson production with vector bosons (VH). The $\ttbar\PH$ production mechanism contributes less than 1\% to the Higgs boson production process and has not been included among the signal processes.

The main background processes, nonresonant $\PQq\PAQq \to \PWp\PWm$ and $\ttbar$+jets, are simulated using the \textsc{MadGraph 5.1.3}~\cite{Alwall:2014hca} and \POWHEG V1~\cite{Alioli:2011as} event generators respectively.  The $\PGg\PGg \to \PWp\PWm$ process is simulated using the \textsc{GG2WW} 3.1 generator~\cite{Binoth:2006mf} and the cross section is scaled to the approximate NLO prediction~\cite{Bonvini:2013jha,Passarino:2013bha}. The tW process is simulated using the \POWHEG V1 generator. Other background processes, such as $\Z/\gamma^{*}\to\tau^{+}\tau^{-}$, $\Z\Z$, $\PW\Z$, $\PW\gamma$, $\PW\gamma^{*}$, tri-bosons (VVV), and $\PW$+jets are generated using \MADGRAPH.

All signal and background generators are interfaced to \textsc{Pythia 6} to simulate the effects of the parton shower, multiple parton interactions, and hadronization.

The default parton distribution function (PDF) sets used are CTEQ6L
\cite{Lai:2010nw} for LO generators and CT10~\cite{Lai:2010vv} for NLO generators.
The $\hww$ process simulation is reweighted so that the \pth{} spectrum and inclusive production cross section closely match the SM calculations that have NNLO+NNLL pQCD accuracy in the description of the Higgs boson inclusive production, in accordance with the LHC Higgs Cross Section Working Group recommendations~\cite{Heinemeyer:2013tqa}.
The reweighting of the \pth{} spectrum is achieved by tuning the \POWHEG generator, as described in detail in Ref.~\cite{Alioli:2010xd}.
Cross sections computed with NLO pQCD accuracy~\cite{Heinemeyer:2013tqa} are used for the background processes.

The samples are processed using a simulation of the CMS detector response, as modeled by \textsc{Geant4}~\cite{Agostinelli:2002hh}.
Minimum bias events are superimposed on the simulated events to emulate the additional pp interactions per bunch crossing (pileup). The events are reweighted to correct for observed differences between data and simulation in the number of pileup events, trigger efficiency, and lepton reconstruction and identification efficiencies~\cite{Chatrchyan:2013iaa}.

For the comparison of the measured unfolded spectrum with the theoretical predictions, two additional MC generators are used for simulating the SM Higgs boson production in the ggH process: \textsc{HRes} 2.3~\cite{deFlorian:2012mx,Grazzini:2013mca} and the second version of the \POWHEG generator (\POWHEG V2)~\cite{Bagnaschi:2011tu}.
\textsc{HRes} is a partonic level MC generator that computes the SM Higgs
boson cross section at NNLO accuracy in pQCD and performs the NNLL
resummation of soft-gluon effects at small \pt. The central predictions of
\textsc{HRes} are obtained including the exact top and bottom quark mass contribution to
the gluon fusion loop, fixing the renormalization and factorization scale central values at a Higgs boson mass of 125\GeV. The cross section normalization is scaled, to take into account electroweak corrections, by a factor of 1.05 and the effects of threshold resummation by a factor of 1.06~\cite{Actis:2008ug,Catani:2003zt}. The upper and lower bounds of the uncertainties are obtained by scaling up and down both the renormalization and the factorization scales by a factor of two.
The \POWHEG V2 generator is a matrix element based generator that provides a NLO description of the ggH process in association with zero jets, taking into account the finite mass of the bottom and top quarks.
The \POWHEG prediction is tuned using the \POWHEG damping factor \textit{hdump} of 104.17\GeV, in order to match the \pth{} spectrum predicted by \textsc{HRes} in the full phase space. This factor reduces the emission of additional jets in the high \pt regime, and enhances the contribution from the Sudakov form factor in the limit of low \pt.
The \POWHEG generator is interfaced to the \textsc{JHUGen} generator version 5.2.5~\cite{Gao:2010qx,Bolognesi:2012mm,Anderson:2013afp} for the decay of the Higgs boson to a W boson pair and interfaced with \PYTHIA 8~\cite{Sjostrand:2007gs} for the simulation of parton shower and hadronization effects.

\section{Analysis strategy}
\label{sec:AnalysisStrategy}

The analysis presented here is based on that  used in the previously published \hwwllvv{}
measurements by CMS~\cite{Chatrchyan:2013iaa}, modified to be inclusive in the number of jets.
This modification significantly reduces the uncertainties related to the modelling of the number of jets produced in association with the Higgs boson because the number of jets is strongly correlated with \pth{}.

Events are selected requiring the presence of two isolated leptons with opposite charge, an electron and a muon, with $\pt > 20 (10)\GeV$ for the leading (subleading) lepton, and with $\abs{\eta}<2.5$ for electrons and $\abs{\eta}<2.4$ for muons. No additional
electron or muon with $\pt>10\GeV$ is allowed. The two leptons are required to
originate from a single primary vertex. Among the vertices identified in the
event, the vertex with the largest $\sum \pt^{2}$, where the sum runs over all
tracks associated with that vertex, is chosen
as the primary vertex.
The invariant mass of the two leptons, \mll{}, is required to be greater than 12\GeV.
A \textit{projected} \MET variable is defined as the component of \ptvecmiss transverse to the nearest lepton if the lepton is situated within the azimuthal angular window of $\pm \pi/2$ from the \ptvecmiss direction, or the \MET itself otherwise~\cite{Chatrchyan:2013iaa}.
Since the \MET resolution is degraded by pileup, the minimum of two projected \MET variables is used: one constructed from all identified particles (full projected \MET), and another constructed from the charged particles only (track projected \MET).
Events must have both \MET and the minimum projected \MET above 20\GeV.
In order to suppress $\Z/\gamma^{*}\to{\tau^{+}\tau^{-}}$ events, the vector
\pt sum of the two leptons, $\pt^{\ell\ell}$, is required to be greater
than 30\GeV and a minimum transverse mass of the lepton plus \MET
vector of 60\GeV is required. The transverse mass is defined as $\mt = \sqrt{\smash[b]{2\pt^{\ell\ell}\ETmiss [ 1-\cos\Delta\phi(\ell\ell, \ptvecmiss) ] }}$, where $\Delta\phi(\ell\ell, \ptvecmiss)$ is the azimuthal angle between the dilepton
momentum and \ptvecmiss.

Events surviving the requirements on leptons are dominantly those where a top quark-antiquark pair is produced and both W bosons, which are part of the top quark decay chain, decay leptonically (dileptonic $\ttbar$).
These events are identified using a b-jet tagging method based on two algorithms: one is the track counting high-efficiency (TCHE)~\cite{Chatrchyan:2012jua}, an algorithm based on the impact parameter of the tracks inside the jet, \ie the distance to the primary vertex at the point of closest approach in the transverse plane; and another is Jet B Probability (JBP), an algorithm that assigns a per track probability of
originating from the primary vertex~\cite{CMS:2013vea}. In addition, soft-muon tagging algorithms are used, which remove events with a nonisolated soft muon, that is likely coming from a b quark decay.

No jet with $\pt>30$\GeV may pass a threshold on the JBP b tagging discriminant corresponding to a b tagging efficiency of 76\% and a mistagging efficiency around 10\%.
No jet with \pt between 15 and 30\GeV may pass a TCHE b tagging discriminant threshold chosen to have a high top quark background rejection efficiency~\cite{Chatrchyan:2013iaa}.
In addition, for events with no reconstructed jets above 30\GeV, a soft-muon veto is applied. Soft muon candidates are defined without isolation requirements and have $\pt>3\GeV$. The efficiency for a $\PQb$ jet with \pt between 15 and 30\GeV to be identified both by the TCHE and soft-muon algorithms is 32\%.

Fiducial phase space requirements are chosen in order to minimize the dependence of the measurements on the underlying model of the Higgs boson properties and its production mechanism.
The exact requirements are determined by considering the two following correlated quantities: the reconstruction efficiency for signal events originating from within the fiducial phase space (fiducial signal efficiency $\epsilon_{\text{fid}}$), and the ratio of the number of reconstructed signal events that are from outside the fiducial phase space (``out-of-fiducial'' signal events) to the number from within the fiducial phase space. The requirement of having a small fraction of out-of-fiducial signal events, while at the same time preserving a high value of the fiducial signal efficiency $\epsilon_{\text{fid}}$, leads to fiducial requirements at the generator level on the low-resolution variables, \ETmiss and \mt, that are looser with respect to those applied in the reconstructed event selection.

The fiducial phase space used for the cross section measurements is defined at the particle level by the requirements given in Table~\ref{table:fid_cuts}. The leptons are defined as Born-level leptons, \ie before the emission of final-state radiation (FSR), and are required not to  originate from leptonic $\tau$ decays. The effect of including FSR is evaluated to be of the order of 5\% in each \pth{} bin.
For the VH signal process the two leptons are required to originate from the \hwwllvv~decays in order to
avoid including leptons coming from the associated W or Z boson.

\begin{table}[ht]
\renewcommand{\arraystretch}{1.1}
\topcaption{Summary of requirements used in the definition of the fiducial phase space. The leptons
are defined at the Born-level.}\label{table:fid_cuts}
\centering
\begin{tabular}{l r}
\hline
Physics quantity & Requirement \\
\hline
Leading lepton \pt & $\pt > 20\GeV$ \\
Subleading lepton \pt & $\pt > 10\GeV$ \\
Pseudorapidity of electrons and muons & \multicolumn{1}{c}{$\abs{\eta} < 2.5$} \\
Invariant mass of the two charged leptons & $\mll > 12\GeV$ \\
Charged lepton pair \pt & $\pt^{\ell\ell} > 30\GeV$ \\
Invariant mass of the leptonic system in the transverse plane & $m_\mathrm{T}^{\ell\ell \nu\nu} > 50\GeV$ \\
\MET & \multicolumn{1}{c}{$\MET>0$} \\
\hline
\end{tabular}

\end{table}

Experimentally, the Higgs boson \pt is reconstructed as the
vector sum of the lepton momenta in the transverse plane and \ptvecmiss:
\begin{equation}
\vec{p}_\mathrm{T}^{\mathrm{H}} = \vec{p}_\mathrm{T}^{{\ell\ell}} + \ptvecmiss .
\end{equation}

Compared to other differential analyses of the Higgs boson $\sigma \, \mathcal{B}$, such as
those in the $\PH \to \Z\Z \to 4\ell$ and $\PH \to \gamma\gamma$ decay channels, this analysis has to cope
with limited resolution due to the \MET entering the \pth{} measurement.
The effect of the limited \MET resolution has two main implications for the
analysis strategy.
The first one is that the choice of the binning in the \pth{} spectrum needs to
take into account the detector resolution. The binning in \pth{} is built
in such a way as to ensure that at least 60\% of the signal events generated
in a given \pth{} bin are also reconstructed in that bin. This procedure
yields the following bin boundaries: {[0, 15]}, {[15, 45]},
{[45, 85]}, {[85, 125]}, {[125, 165]},
and {[165, $\infty$]}\GeV.
The second implication is that migrations of events across bins are significant.

The signal yield is extracted in each $\pth$ bin with a template fit to a two
dimensional distribution of $\mll$ and $\mt$. These two observables are chosen for the template
fit because they are weakly correlated with \pth{}. The level of correlation is checked using simulation.

\section{Background estimation}
\label{sec:selections}

The signal extraction procedure requires the determination of the normalization and ($\mll,~\mt$) shape for each background source.
After the event selection is applied, one of the dominant contributions to the background processes arises from the top quark production, including the dileptonic $\ttbar$ and tW processes.
The top quark background is divided into two categories with different jet multiplicity: the first category requires events without jets with \pt above 30\GeV and the second one requires at least one jet with $\pt>30\GeV$.
For the estimation of the top quark background in the first category, the same estimate from control samples in data as in Ref.~\cite{Chatrchyan:2013iaa} is used.
The contribution of the background in the second category is estimated independently in each \pth{} bin, by normalizing it in a control region defined by requiring at least one jet with a JBP b tagging discriminator value above a given threshold, chosen to have a pure control region enriched in $\PQb$ jets.
In addition, the quality of the Monte Carlo description of ($\mll,~\mt$) kinematics is verified for this background by looking at the shapes of these variables in the $\PQb$ jets enriched control region and is found to be satisfactory.

The nonresonant $\PQq\PAQq \to \PWp\PWm$ is
determined independently in each \pth{} bin.
The shape of the ($\mll,~\mt$) distribution for this background is taken from
the simulation, and its normalization in each \pth{} bin is obtained from the
template fit of the ($\mll,~\mt$) distribution, together with the signal
yield.
Approximately 5\% of the \wwllvv{} originates from a gluon-gluon initial state
via a quark box diagram. This background is treated separately and both normalization and shape are taken from simulation.

Backgrounds containing one or two misidentified leptons are estimated from events
selected with relaxed lepton quality criteria, using the techniques
described in Ref.~\cite{Chatrchyan:2013iaa}.

The \dytt{} background process is estimated using \dymm{} events selected in
data, in which the muons are replaced with simulated $\tau$ decays, thus
providing a more accurate description of the experimental conditions than the full simulation~\cite{Chatrchyan:2013iaa}. The \textsc{tauola} package~\cite{Jadach:1990mz} is used in the simulation of $\tau$ decays to account for $\tau$-polarization effects.

Contributions from $\PW\gamma^{*}$ and $\PW\gamma$ production processes are estimated partly from simulated samples. The $\PW\gamma^{*}$ cross section is measured from data and the discriminant variables used in the signal extraction for the $\PW\gamma$ process are obtained from data as explained in Ref.~\cite{Chatrchyan:2013iaa}. The shape of the discriminant variables for the $\PW\gamma^{*}$ process and the $\PW\gamma$ cross section are taken from simulation.

A summary of the processes used to estimate backgrounds is reported in Table~\ref{table:bkg_estimation}. The normalization and shape of the backgrounds are estimated using data control samples whenever possible.
The remaining minor background contributions are estimated using simulation. The yield of each background process after the analysis requirements is given in Section~\ref{sec:signalExtraction}.

\begin{table}[ht]
\topcaption{Summary of the processes used to estimate backgrounds in cases where data events are used to estimate either the normalization or the shape of the discriminant variable. A brief description of the control/template samples is given.}\label{table:bkg_estimation}
\centering
\begin{tabular}{l lll}\hline
{Process} & {Normalization} & {Shape} & {Control/template sample} \rule{0pt}{2.3ex} \\
\hline
$\PW\PW$ & data & simulation & events at high \mll~and \mt \rule{0pt}{2.3ex} \\
Top & data & simulation & \begin{tabular}[c]{@{}c@{}}${\ge}2$~jets with at least one\rule{0pt}{2.3ex} \\ passing $\PQb$ tagging criteria\end{tabular}  \\
$\PW$+jets & data & data & events with loosely identified leptons\rule{0pt}{2.3ex} \\
$\PW\gamma$ & simulation & data & events with an identified $\gamma$\rule{0pt}{2.3ex} \\
$\PW\gamma^*$ & data & simulation & $\PW\gamma^* \to 3\mu$ sample\rule{0pt}{2.3ex} \\
Z$/\gamma^* \to \tau\tau$ & data & data & $\tau$ embedded sample\rule{0pt}{2.3ex} \rule[-1.0ex]{0pt}{0pt}\\
\hline
\end{tabular}

\end{table}

\section{Systematic uncertainties}
\label{sec:systematicUncertainties}

Systematic uncertainties in this analysis arise from three sources: background predictions, experimental measurements, and
theoretical uncertainties.

The estimates of most of the systematic uncertainties use the same methods as the published $\hwwllvv$ analysis \cite{Chatrchyan:2013iaa}.
One notable difference is in the uncertainties related to the prediction of
the contributions from $\ttbar$ and tW processes. The shapes of these
backgrounds are corrected for different b tagging efficiency in data and MC
simulation, and the normalization is taken from data in a top quark enriched control region independently in each \pth{} bin, as explained in Section~\ref{sec:selections}.
The uncertainties related to this procedure arise from the sample size in the control regions for each \pth{} bin, and are embedded in the scale factors used to extrapolate the top quark background normalization from the control region to the signal region. They vary from $20\%$ to $50\%$ depending on the \pth{} bin.

This analysis takes into account the theoretical uncertainties that affect the normalization and shape of all backgrounds and the signal distribution shape. These uncertainties arise from missing higher-order corrections in pQCD and PDF uncertainties, and are predicted using MC simulations.
The effect due to the variations in the choice of PDFs and the value of the QCD coupling constant is considered following the PDF4LHC~\cite{Alekhin:2011sk,Botje:2011sn} prescription, using CT10, NNPDF2.1~\cite{Ball:2011mu} and MSTW2008~\cite{Martin:2009iq} PDF sets.

The uncertainties in the signal yield associated with the uncertainty in the ($\mll,~\mt$) shapes due to the missing higher-order corrections
are evaluated independently by varying up and down the factorization and
renormalization scales by a factor of two, and then using the Stewart-Tackman
formulae~\cite{Stewart:2011cf}. Due to the presence of the b-veto, the uncertainty on jet multiplicity must be evaluated. However, this uncertainty is diluted since the b-veto efficiency is weakly dependent on the number of jets in the event.

Since the shapes of the $\PW\PW$ background templates used in the fit are taken from MC simulation, a corresponding shape uncertainty must be accounted for. This uncertainty is estimated in each bin of \pth{} from the comparisons of the two estimates obtained using the sample produced with \textsc{MadGraph 5.1.3}, and another sample produced using \textsc{mc@nlo 4.0}~\cite{Frixione:2010wd}.
These uncertainties include shape differences originating
from the renormalization and factorization scale choice. The scale dependence is estimated with \textsc{mc@nlo}.

A summary of the main sources of systematic uncertainty and the corresponding estimate is reported in Table~\ref{tab:Systematics}.

The systematic uncertainties related to the unfolding procedure are described separately in Section~\ref{sec:unfoldingAndSyst}.

\begin{table}[ht]
\renewcommand{\arraystretch}{1.1}
  \centering
  \topcaption{Main sources of systematic uncertainties and their estimate. The
  first category reports the uncertainties in the normalization of background
  contributions. The experimental and theoretical uncertainties refer to the
  effect on signal yields. A range is specified if the uncertainty varies
  across the $\pth$ bins.}
  \label{tab:Systematics}
  \begin{tabular}{ll}
  \multicolumn{2}{c} {{Uncertainties in backgrounds contributions}} \\
  \hline
  {Source}  & {Uncertainty} \\
  \hline
  $\ttbar$, tW      & 20--50\% \\
  $\PW$+jets              & 40\% \\
  $\PW\Z$, $\Z\Z$              & \x4\% \\
  $\PW\gamma^{(*)}$  & 30\% \\
  \hline
  \multicolumn{2}{c} {{Effect of the experimental uncertainties on the signal and background yields}}\rule{0pt}{5.0ex}\\
  \hline
  {Source} & {Uncertainty}\\
  \hline
  Integrated luminosity        & 2.6\% \\
  Trigger efficiency           & 1--2\% \\
  Lepton reconstruction and identification & 3--4\%\\
  Lepton energy scale          & 2--4\% \\
  $\ETmiss$ modelling          & \x2\% \\
  Jet energy scale             & 10\% \\
  Pileup multiplicity          & \x2\% \\
  $\PQb$ mistag modelling	       & \x3\% \\	
  \hline
  \multicolumn{2}{c}{{Effect of the theoretical uncertainties on signal yield}}\rule{0pt}{5.0ex}\\
  \hline
  {Source} & {Uncertainty} \\
  \hline
  $\PQb$ jet veto scale factor              & 1--2\% \\
  PDF                                  & \x1\% \\
  $\PW\PW$ background shape                  & \x1\% \\
  \hline
  \end{tabular}
\end{table}

\section{Signal extraction}
\label{sec:signalExtraction}

The signal, including ggH, VBF, and VH production mechanisms, is extracted in each bin of \pth{} by performing a binned maximum likelihood fit simultaneously in all \pth{} bins to a two-dimensional template for signals and backgrounds in the \mll--\mt{} plane.
Six different signal strength parameters are extracted from the fit, one for each \pth~bin. The relative contributions of the different Higgs production mechanisms in the signal template are taken to be the same as in the SM.
The systematic uncertainty sources are considered as nuisance parameters in the fit.

Because of detector resolution effects, some of the reconstructed $\hww$ signal events might originate from outside the fiducial phase space.
These out-of-fiducial signal events cannot be precisely handled by the unfolding procedure and must be subtracted from the measured spectrum. The \pth~distribution of the out-of-fiducial signal events is taken from simulation, and each bin is multiplied by the corresponding measured signal strength before performing the subtraction.

\begin{figure}[!htbp]
\centering
{
\includegraphics[width=0.35\textwidth]{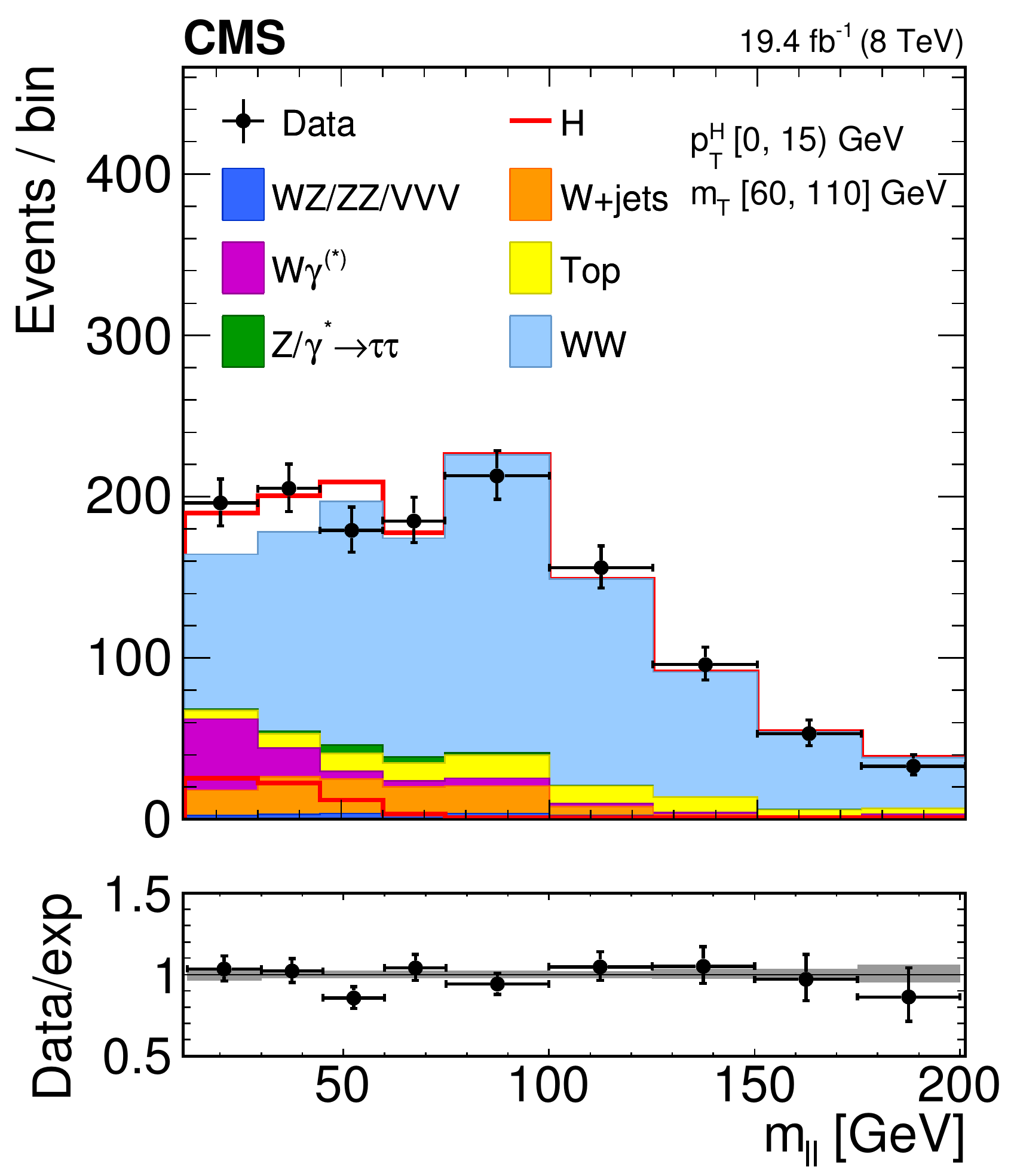}
}
{
\includegraphics[width=0.35\textwidth]{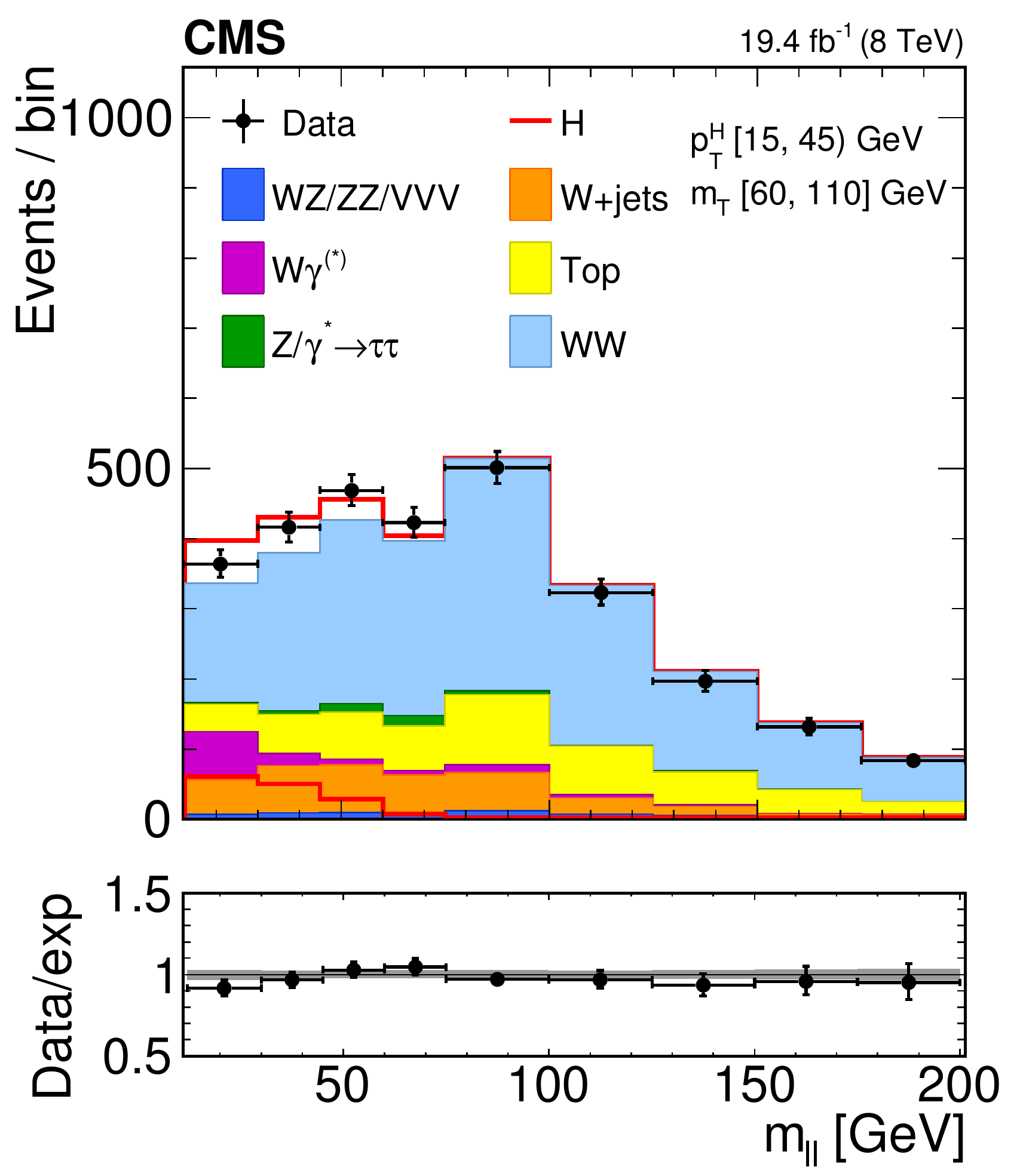}
}
\\
{
\includegraphics[width=0.35\textwidth]{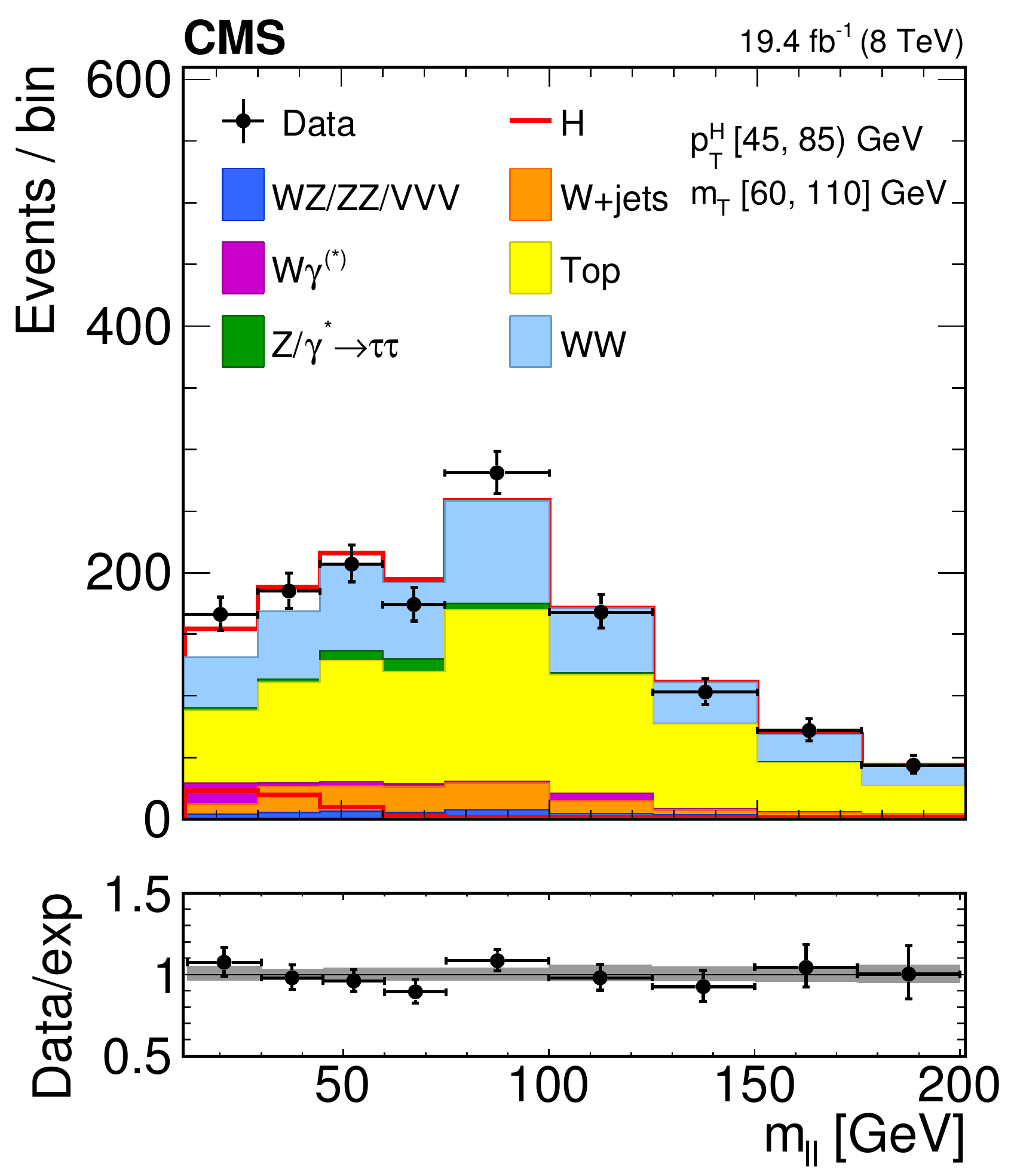}
}
{
\includegraphics[width=0.35\textwidth]{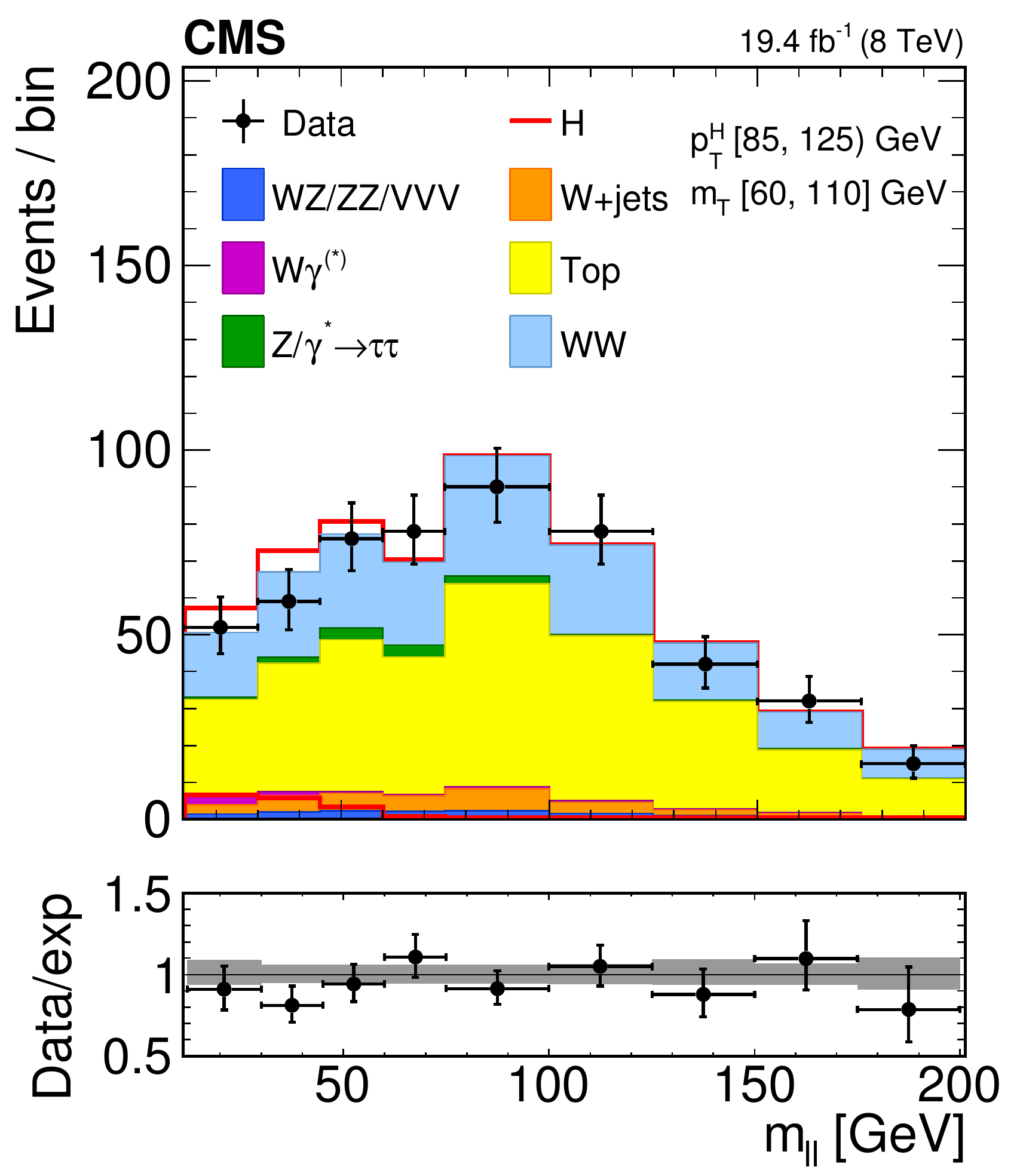}
}
\\
{
\includegraphics[width=0.35\textwidth]{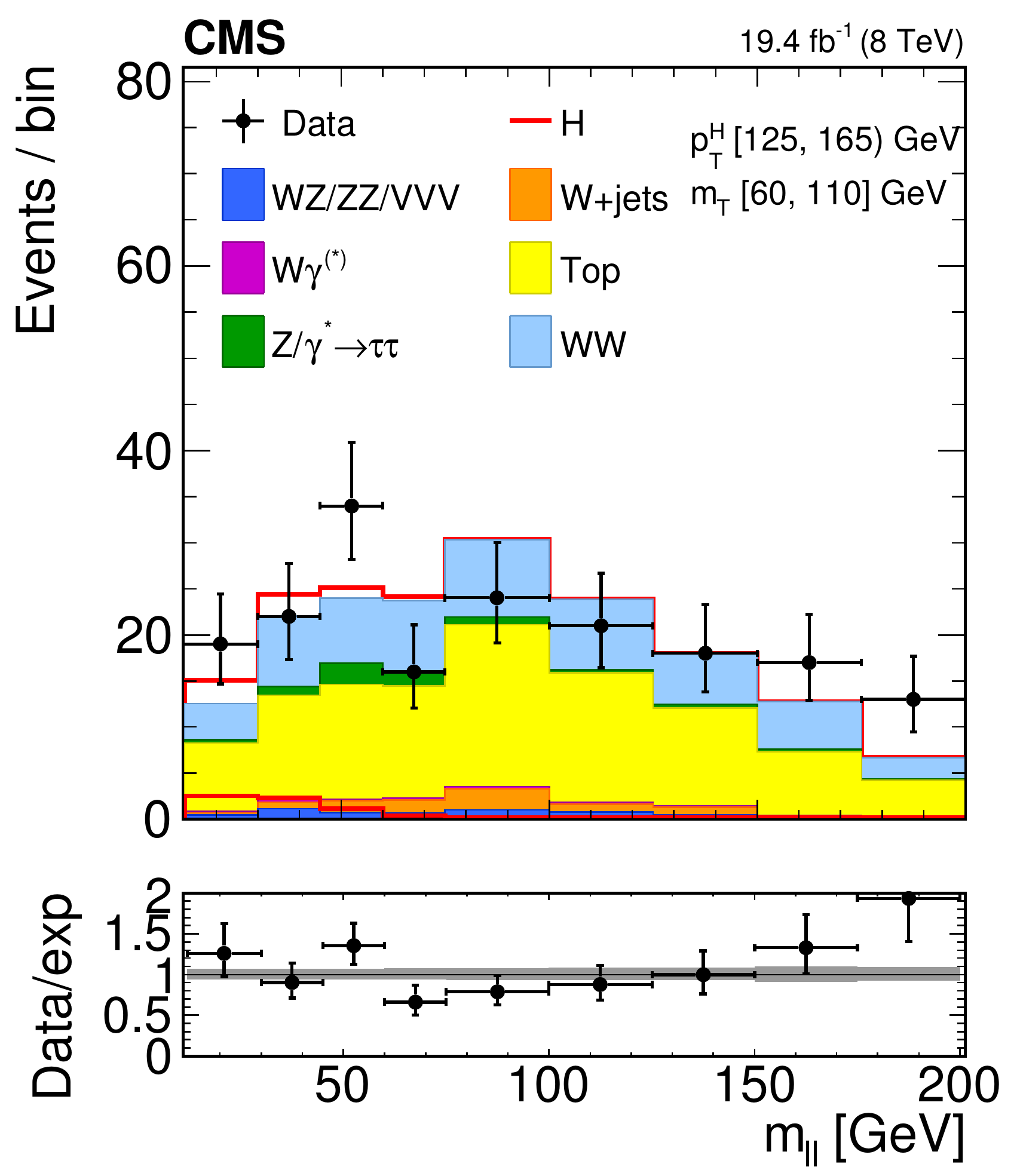}
}
{
\includegraphics[width=0.35\textwidth]{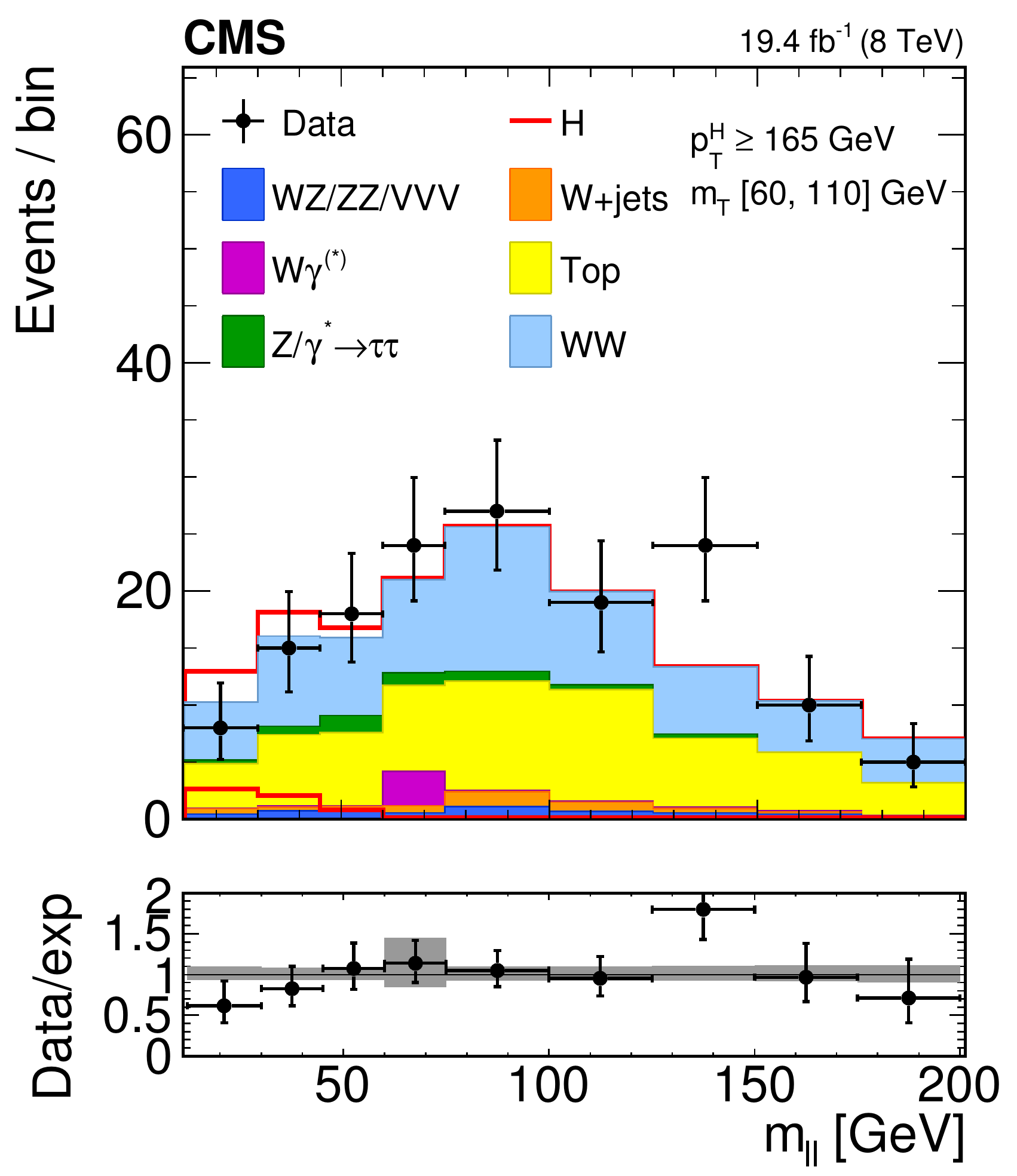}
}
\caption{Distributions of the \mll~variable in each of the six \pth{} bins. Background normalizations correspond to the values obtained from the fit. Signal normalization is fixed to the SM expectation. The distributions are shown in an \mt{}~window of [60,110]\GeV in order to emphasize the Higgs boson (H) signal. The signal contribution is shown both stacked on top of the background and superimposed on it. Ratios of the expected and observed event yields in individual bins are shown in the panels below the plots. The uncertainty band shown in the ratio plot corresponds to the envelope of systematic uncertainties after performing the fit to the data.}\label{fig:mllSignalRegion}
\end{figure}

\begin{figure}[!htbp]
\centering
{
\includegraphics[width=0.35\textwidth]{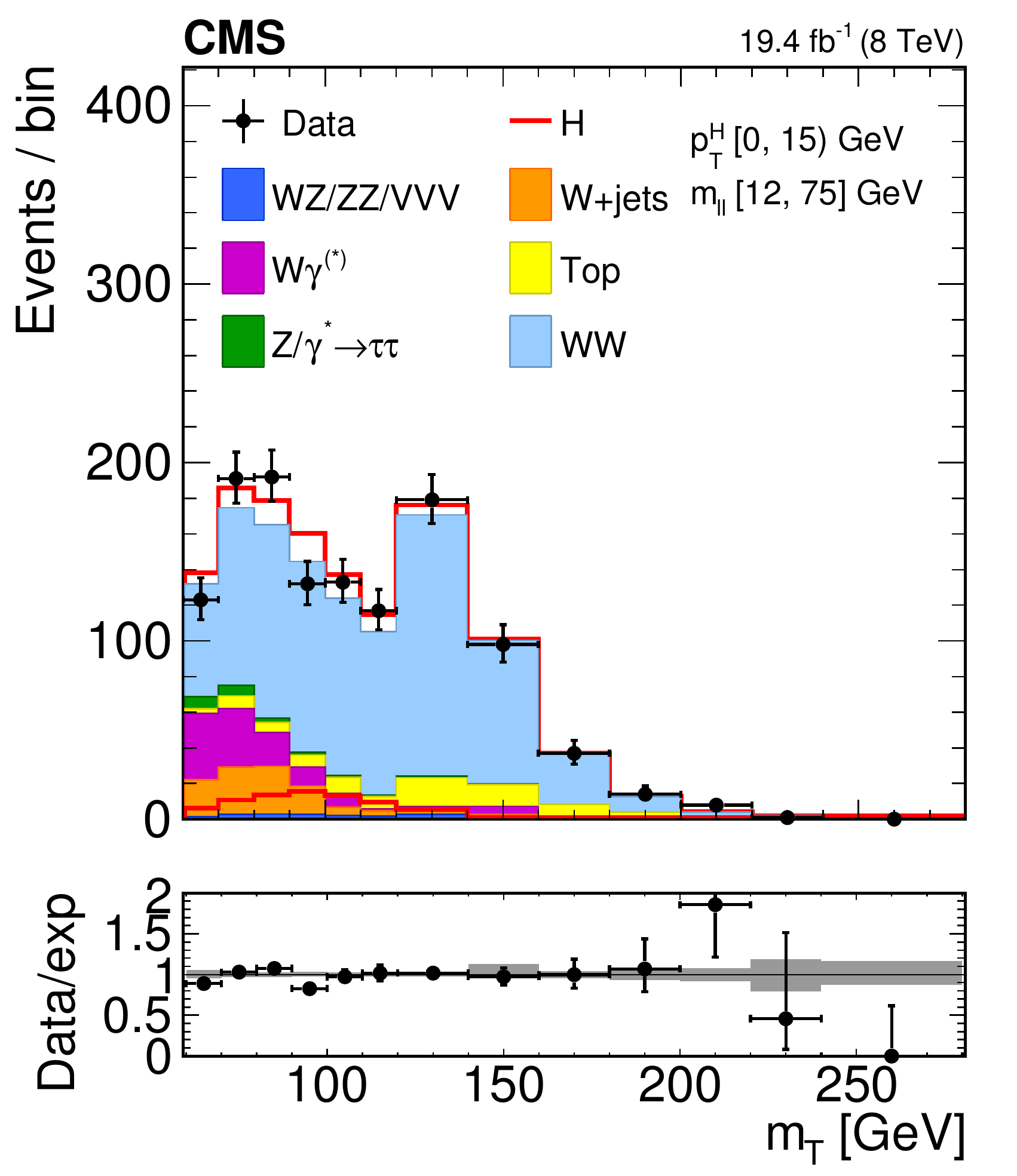}
}
{
\includegraphics[width=0.35\textwidth]{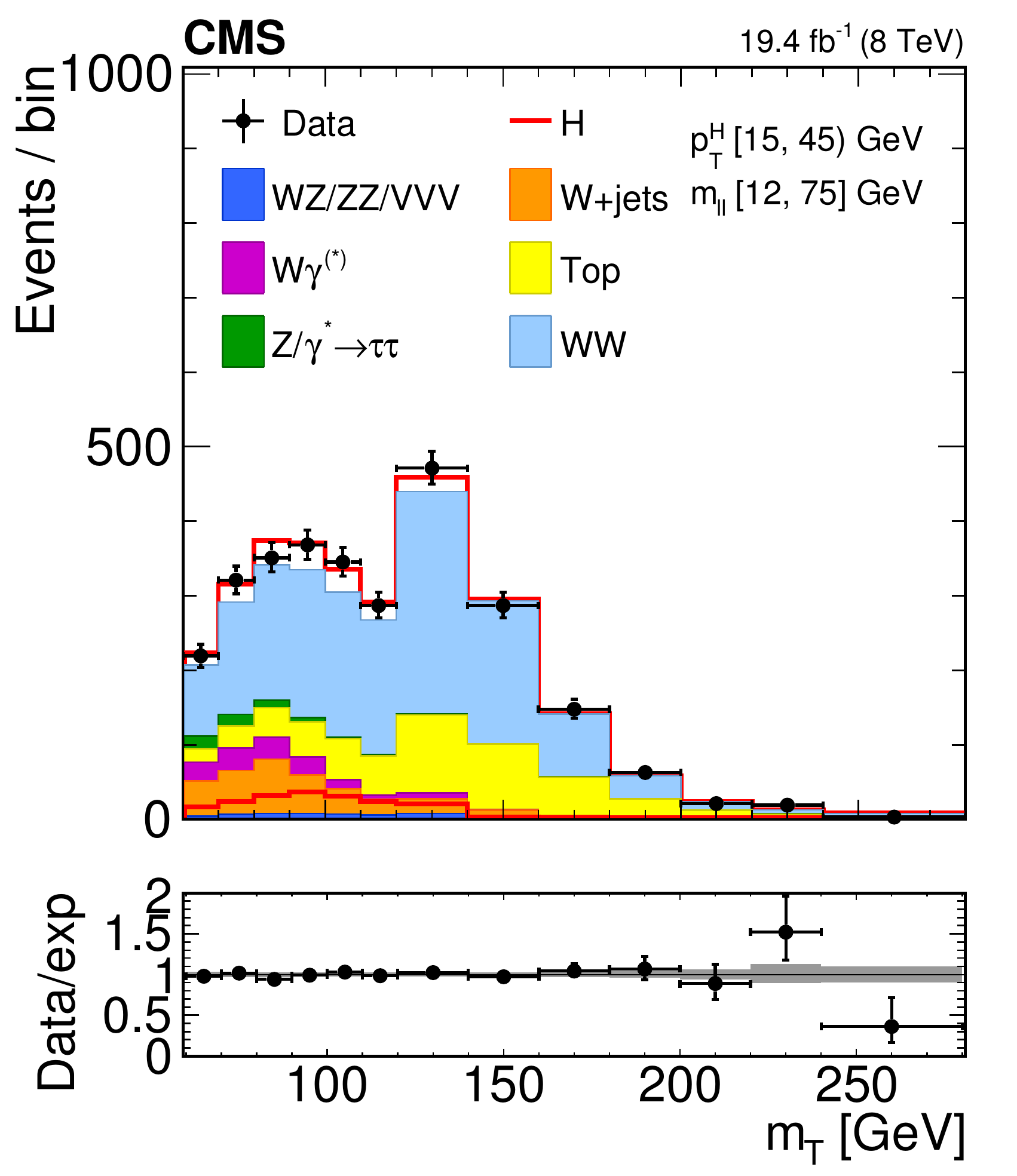}
}
\\
{
\includegraphics[width=0.35\textwidth]{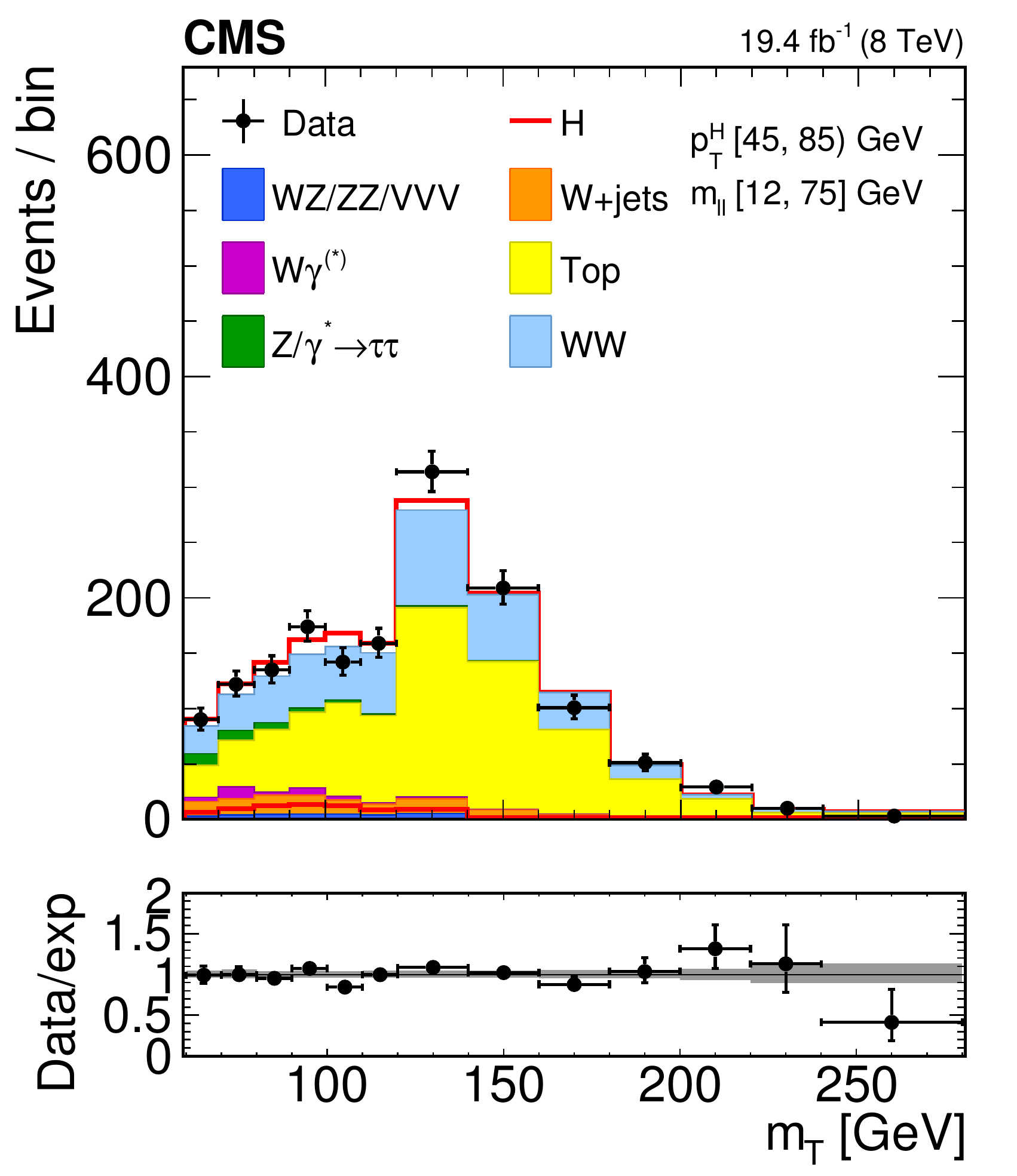}
}
{
\includegraphics[width=0.35\textwidth]{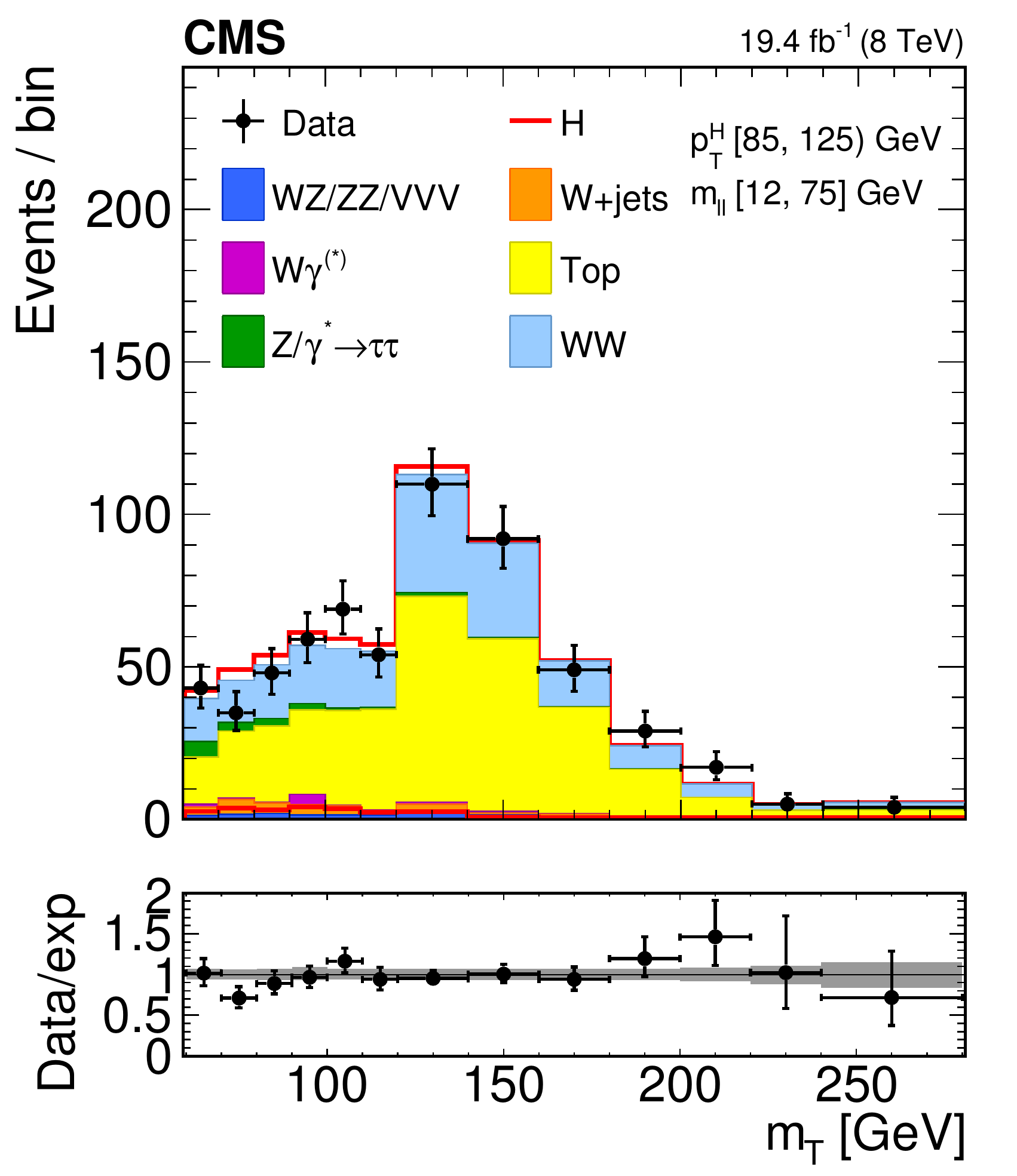}
}
\\
{
\includegraphics[width=0.35\textwidth]{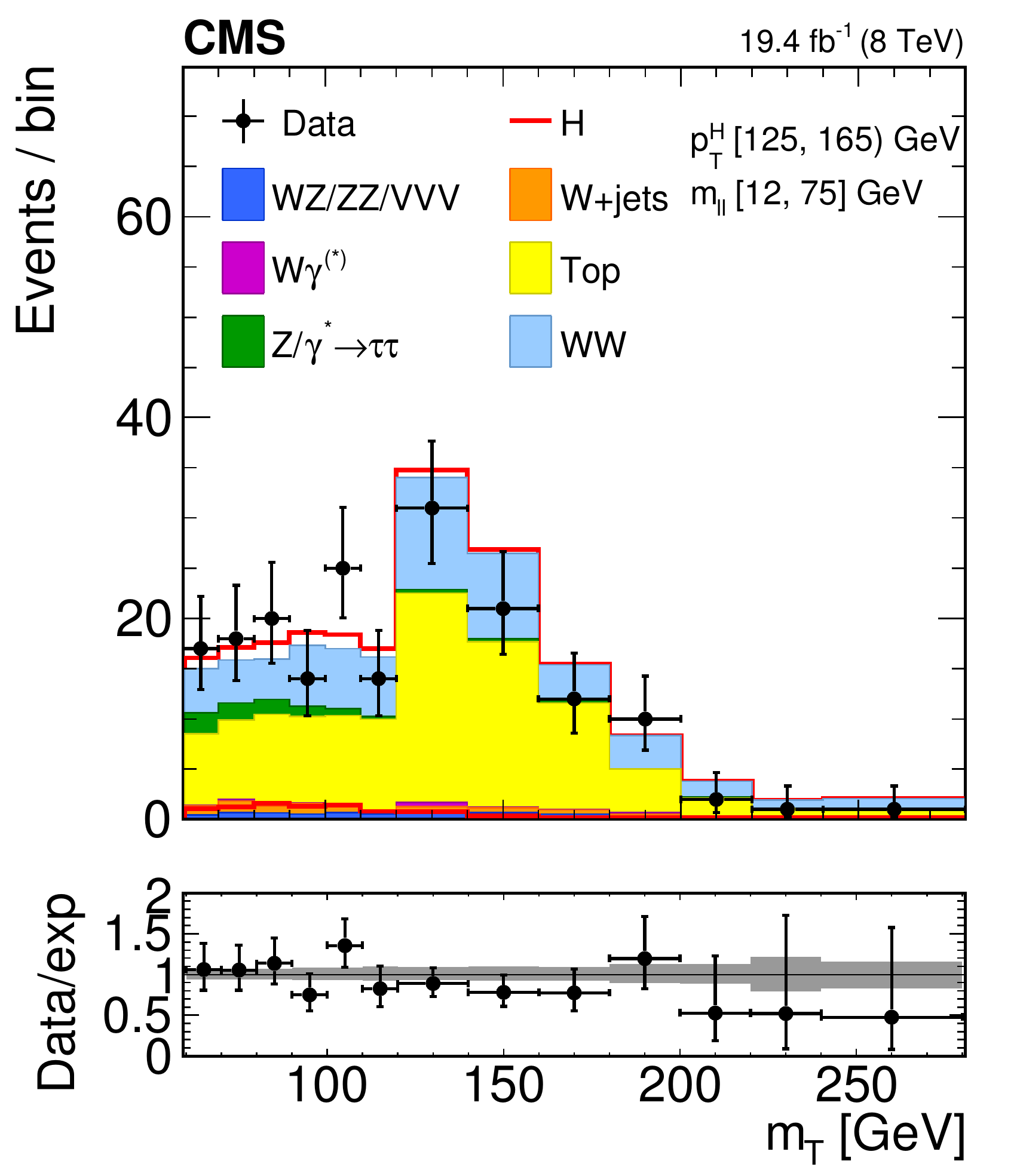}
}
{
\includegraphics[width=0.35\textwidth]{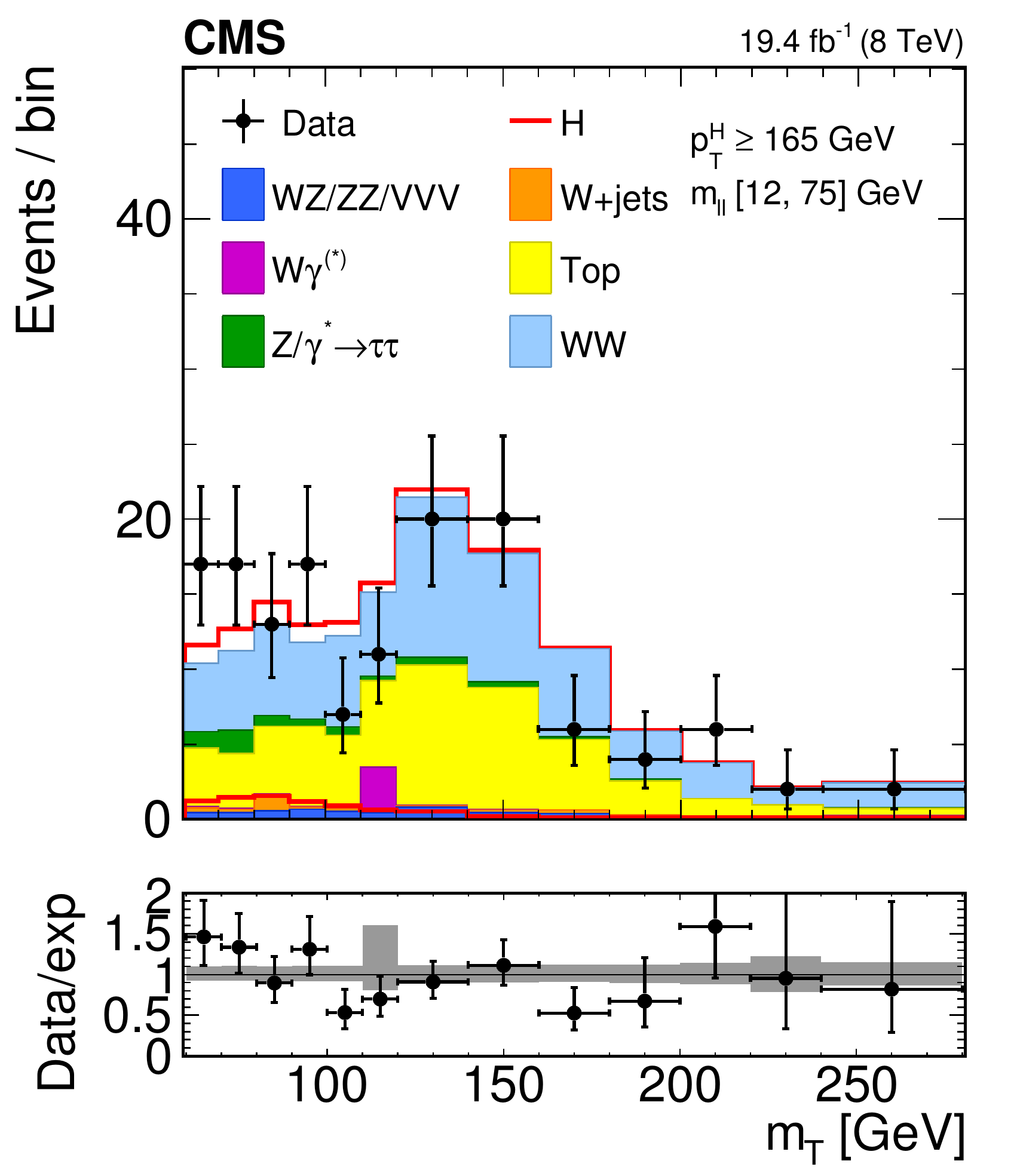}
}
\caption{Distributions of the \mt~variable in each of the six \pth{} bins. Background normalizations correspond to the values obtained from the fit. Signal normalization is fixed to the SM expectation. The distributions are shown in an \mll{}~window of [12,75]\GeV in order to emphasize the Higgs boson (H) signal. The signal contribution is shown both stacked on top of the background and superimposed on it. Ratios of the expected and observed event yields in individual bins are shown in the panels below the plots. The uncertainty band shown in the ratio plot corresponds to the envelope of systematic uncertainties after performing the fit to the data.}\label{fig:mTSignalRegion}
\end{figure}

A comparison of data and background prediction is shown in
Fig.~\ref{fig:mllSignalRegion}, where the \mll{} distribution is shown for each of the six \pth{} bins. Distributions correspond to the \mt{} window of [60, 110]\GeV, in order to emphasize the Higgs boson signal~\cite{Chatrchyan:2013iaa}. The corresponding \mt{} distributions are shown in Fig.~\ref{fig:mTSignalRegion} for events in an \mll{} window of [12, 75]\GeV.

The signal prediction and background estimates after the analysis selection are reported in Table~\ref{table:yields}. Background normalizations correspond to the values obtained from the fit.

\begin{table}[htbp]
\renewcommand{\arraystretch}{1.1}
\centering
 {
  \topcaption{Signal prediction, background estimates and observed number of events in data are shown in each \pth{} bin for the signal after applying the analysis selection requirements. The total uncertainty on the number of events is reported. For signal processes, the yield related to the ggH are shown, separated with respect to the contribution of the other production mechanisms (XH=VBF+VH). The $\PW\PW$ process includes both quark and gluon induced contribution, while the Top process takes into account both $\ttbar$ and tW. }\label{table:yields}
  \resizebox{\textwidth}{!}{
\begin{tabular} {l c c c c c c}
  \hline
$\pt^{\mathrm{H}} [\GeVns{}] $	&	0--15	&	15--45	&	45--85	&	85--125	&	125--165	&	165--$\infty$ \\ 	
\hline
ggH	&	$73\pm3\x$	&	$175\pm5\xx$	&                $59\pm3\x$	&                $15\pm2\x$	&                $5.1\pm1.5$	&                $4.9\pm1.4$	\\
XH=VBF+VH	&	$4\pm2$ 	&	$15\pm4\x$ 	&		 $16\pm4\x$	&	         $8\pm2$ 	&		 $3.8\pm1.1$ 	&		 $3.0\pm0.8$    \\	
Out-of-fiducial & $9.2\pm0.5$   &       $19.9\pm0.7\x$    &      $11.4\pm0.6\x$    &    $4.4\pm0.3$   &     $1.6\pm0.2$   &   $2.4\pm0.2$ \\
Data 	&	2182	 	&         5305	 	&	         3042	 	& 	          1263	 	&	         431	 	& 	          343	 	\\
Total background &  	 $2124\pm128\x$	 &     $5170\pm321\x$	 &       $2947\pm293\x$	 &            $1266\pm175\x$	 &         $420\pm80\x$	 &              $336\pm74\x$	 \\
$\PW\PW$ 	& 		$1616\pm107\x$	 &	 $3172\pm249\x$	 &	     $865\pm217$	 &	     $421\pm120$	 &	     $125\pm60\x$	 &		     $161\pm54\x$	 \\
Top 	&	$184\pm38\x$	&	                $1199\pm165\x$	&	                $1741\pm192\x$	&	                $735\pm125$	&	                $243\pm51\x$	& 	        $139\pm49\x$	\\
$\PW$+jets 	& $134\pm5\xx$ 	&	         $455\pm10\x$ 	&	         $174\pm6\xx$ 	&	         $48\pm4\x$ 	&	         $14\pm3\x$ 	&	         $9\pm3$ 	\\
$\PW\Z$+$\Z\Z$+VVV & $34\pm4\x$ 	 &	$107\pm10\x$ 	&                $71\pm7\x$ 	&	         $29\pm5\x$ 	&                $14\pm3\x$ 	&     $13\pm4\x$ 	\\
\dytt 	&	$23\pm3\x$ 	&         $67\pm5\x$ 	&         $47\pm4\x$ 	&         $22\pm3\x$ 	&         $12\pm2\x$ 	&         $10\pm2\x$ 	\\
$\PW\gamma^{(*)}$	& $132\pm49\x$     &             $170\pm58\x$    &              $48\pm30$ &                  $12\pm9\x$ &                  $3\pm3$ &                 $\x5\pm10$ \\
\hline
  \end{tabular}
  }
  }

\end{table}

\begin{figure}[!ht]
\centering
\includegraphics[width=0.6\textwidth]{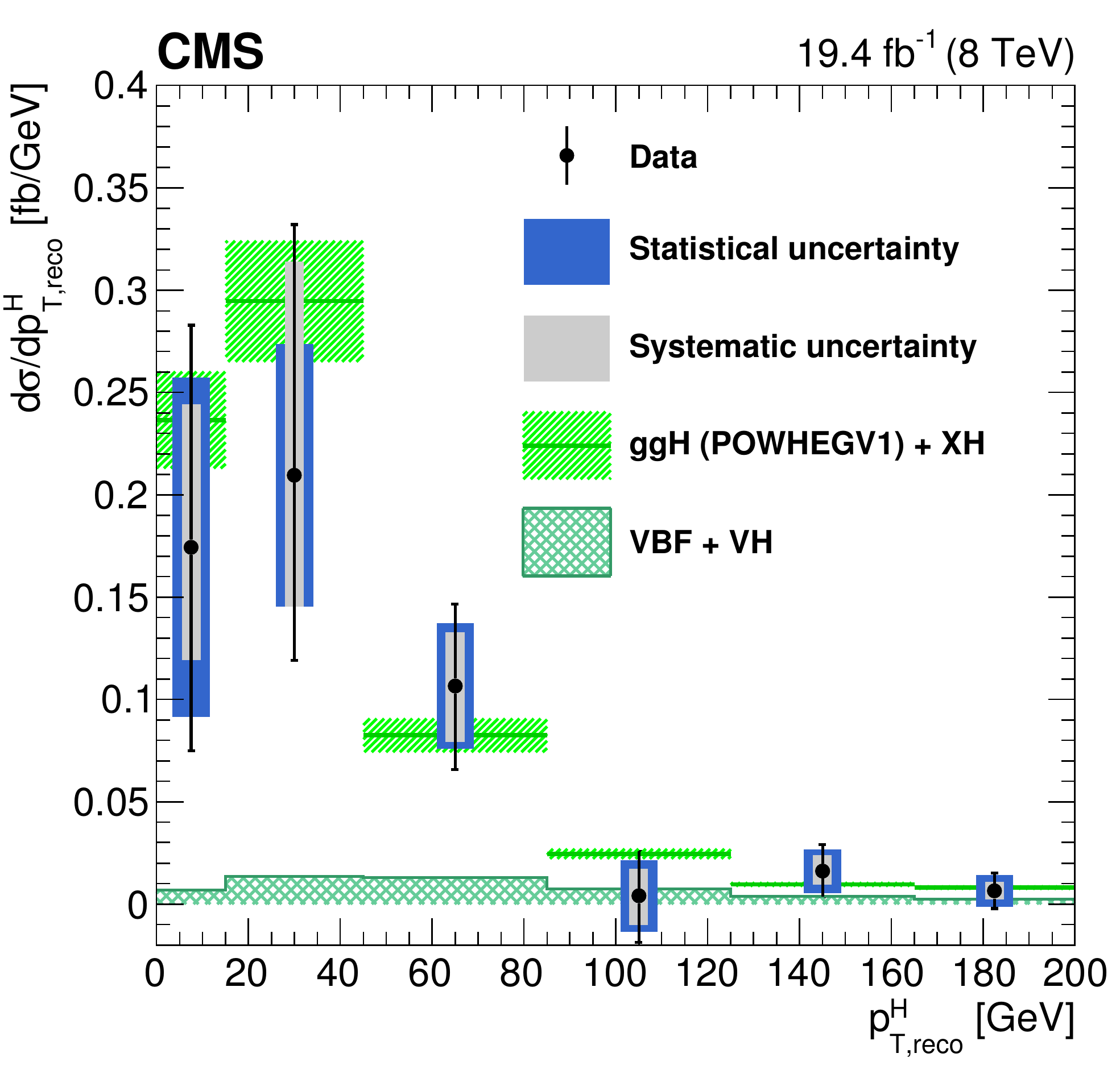}
\caption{Differential Higgs boson production cross section as a function of the reconstructed \pth{}, before applying the unfolding procedure. Data values after the background subtraction are shown together with the statistical and the systematic uncertainties, determined propagating the sources of uncertainty through the fit procedure. The line and dashed area represent the SM theoretical estimates in which the acceptance of the dominant ggH contribution is modelled by \POWHEG V1. The sub-dominant component of the signal is denoted as XH=VBF+VH, and is shown with the cross filled area separately.}\label{fig:pre_unfolding}
\end{figure}

The spectrum shown in Fig. \ref{fig:pre_unfolding} is obtained after having performed the fit and after the subtraction of the out-of-fiducial signal events, but before undergoing the unfolding procedure. The theoretical distribution after the detector simulation and event reconstruction is also shown for comparison.

\section{Unfolding and treatment of systematic uncertainties}
\label{sec:unfoldingAndSyst}

To facilitate comparisons with theoretical predictions or other experimental results, the signal
extracted performing the fit has to be corrected for detector resolution and
efficiency effects and for the efficiency of the selection defined in the
analysis.
An unfolding procedure is used relying on the \textsc{RooUnfold} package
\cite{Adye:2011gm}, which provides the tools to run various unfolding
algorithms.

For every variable of interest, simulated samples are used to compare the
distribution of that variable before and after the simulated events are
processed through CMS detector simulation and events reconstruction.
The detector response matrix $M$ is built according to the following equation:
\begin{equation}\label{eq:resp_matrix}
R_{i}^{\mathrm{MC}} = \sum_{j=1}^{n} M_{ij}T_{j}^{\mathrm{MC}} ,
\end{equation}
where $T^{\mathrm{MC}}$ and $R^{\mathrm{MC}}$ are two $n$-dimensional vectors
representing the distribution before and after event processing through CMS
simulation and reconstruction, respectively. The dimension $n$ of the two vectors corresponds
to the number of bins in the distributions, equal to six in this analysis.
The response matrix $M$ includes all the effects related to the detector and analysis selection that affect the $R^{\mathrm{MC}}$ distribution.
To avoid the large variance and strong negative correlation between the neighbouring bins~\cite{Cowan:2002in}, the unfolding procedure in this analysis relies on the singular value decomposition~\cite{Hocker:1995kb} method based on the Tikhonov regularization
function.
The regularization parameter is chosen to obtain results that are robust against numerical instabilities and statistical fluctuations, following the prescription described in Ref.~\cite{Hocker:1995kb}.
It has been verified using a large number of simulated pseudo-experiments that the coverage of the unfolded uncertainties obtained with this procedure is as expected.

\begin{figure}[!ht]
\centering
\includegraphics[width=0.45\textwidth]{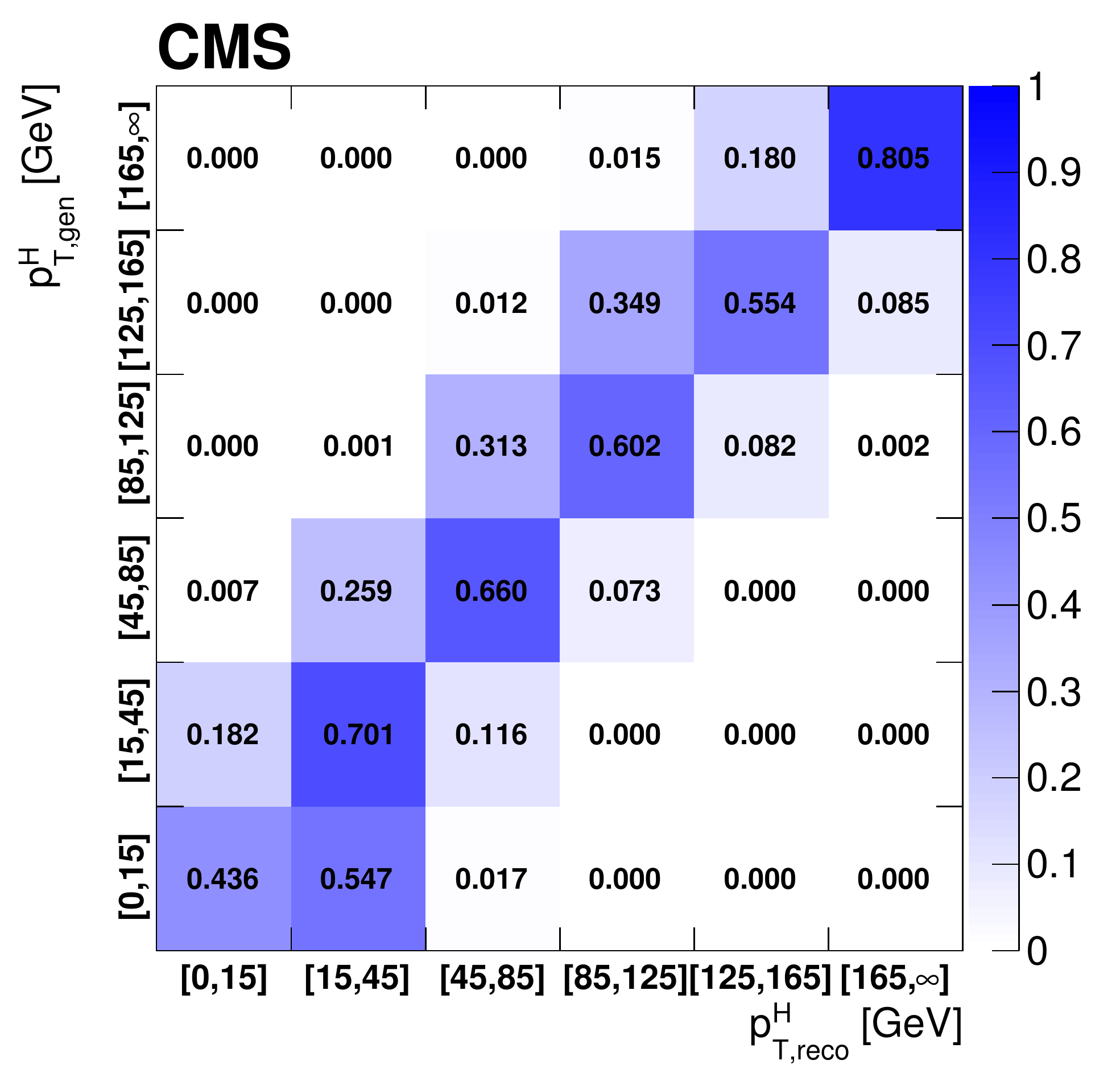}
\hspace{0.5cm}
\includegraphics[width=0.45\textwidth]{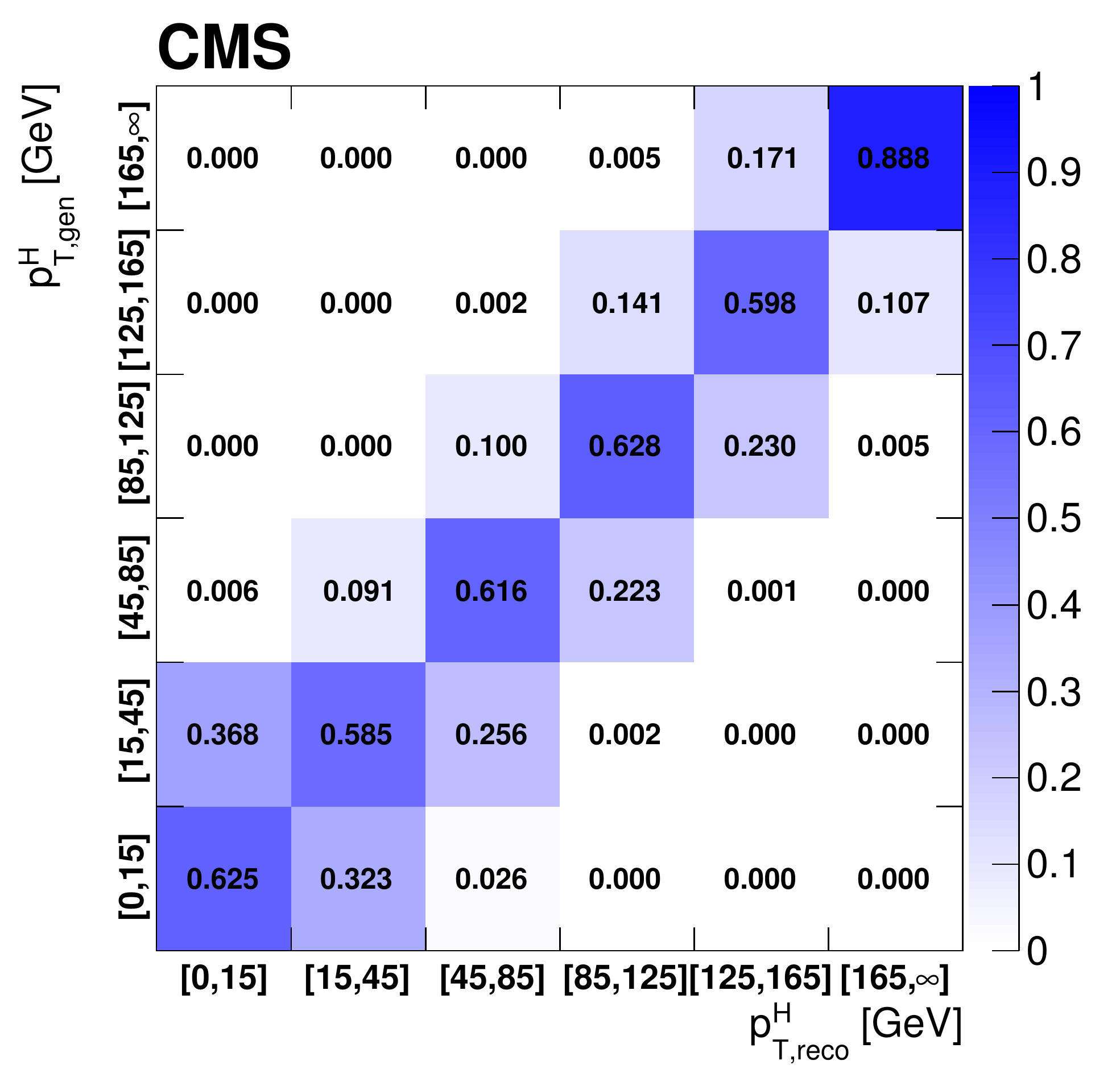}
\caption{Response matrix normalized by row (left) and by column (right) including all signal processes. The matrices are normalized either by row (left) or by column (right) in order to show the purity or stability respectively in diagonal bins.}\label{fig:matrix}
\end{figure}

The response matrix is built as a two-dimensional histogram, with the
generator-level \pth{} on the $y$ axis and the same variable after the reconstruction on the $x$ axis, using the same binning for both
distributions.
The resulting detector response matrix, including all signal sources and normalized by row, is shown in Fig.~\ref{fig:matrix}(left).
The diagonal bins correspond to the purity $P$, defined as the ratio of the
number of events generated and reconstructed in a given bin, to the number of events generated in that bin. The same matrix, normalized by column, is shown in Fig.~\ref{fig:matrix}(right). In this case the diagonal bins correspond to the stability $S$, defined as the ratio of the number of events generated and reconstructed in a given bin, and the number of events reconstructed in that bin. The $P$ and $S$ parameters provide an estimate of the \pth{} resolution and migration effects.
The main source of bin migrations effects in the response matrix is the limited resolution in the measurement of \MET.

Several closure tests are performed in order to validate the unfolding
procedure. To estimate the uncertainty in the unfolding procedure due to the
particular model adopted for building the response matrix, two independent
gluon fusion samples are used, corresponding to two different generators:
\POWHEG V1 and \textsc{JHUGen} generators, both interfaced to \textsc{Pythia 6.4}.
The \textsc{JHUGen} generator sample is used to build the response matrix while the
\POWHEG V1 sample is used for the measured and the MC distributions at
generator level. The result of this test shows good agreement between the unfolded and the distribution from MC simulation.

An important aspect of this analysis is the treatment of the systematic
uncertainties and the error propagation through the unfolding procedure.
The sources of uncertainty are divided into three categories, depending
on whether the uncertainty affects only the signal yield (type A), both the signal
yield and the response matrix (type B), or only the response matrix (type C).
These three classes propagate differently through the unfolding procedure.

Type A uncertainties are extracted directly from the fit in the form of a covariance
matrix, which is passed to the unfolding tool as the covariance
matrix of the measured distribution. The nuisance parameters belonging to this category
are the background shape and normalization uncertainties.
To extract the effect of type A uncertainties a dedicated fit is performed,
fixing to constant all the nuisance parameters in the model, but type A nuisance parameters.

The nuisance parameters falling in the type B class are:
\begin{itemize}
\item the $\PQb$ veto scale factor. It affects the signal and background templates
by varying the number of events with jets that enter the selection. It also
affects the response matrix because the reconstructed spectrum is harder or softer depending on the number of jets, which in turn depends on the veto.
\item the lepton efficiency scale factor. It affects the signal and background
template shape and normalization. It affects the response matrix by varying
the reconstructed spectrum;
\item the \MET scale and resolution, which have an effect similar to the above;
\item lepton scale and resolution. The effect is similar to the above;
\item jet energy scale. It affects the signal and background template shape
and normalization. It also affects the response matrix because, by varying the
fraction of events with jets, the $\PQb$ veto can reject more or fewer events, thus
making the reconstructed spectrum harder or softer.
\end{itemize}
The effect of each type B uncertainty is evaluated separately,
since each one changes the response matrix in a different way.
In order to evaluate their effect on the signal strengths parameters, two additional fits are
performed, each time fixing  the nuisance parameter value to ${\pm} 1$  standard
deviation with respect
to its nominal value. The results of the fits are then compared to the results of the full fit obtained by floating  all the nuisance parameters, thus
determining the relative uncertainty on the signal strengths due to each
nuisance parameter.
Using these uncertainties, the measured spectra for each type B
source are built.
The effects are propagated through the unfolding by building the corresponding variations of the response matrix and unfolding the
measured spectra with the appropriate matrix.

Type C uncertainties are related to the underlying assumption on the Higgs boson production mechanism used to extract the fiducial cross sections. These are evaluated using an alternative shape for the true distribution at generator level. Since the reconstructed spectrum observed in data is consistent with a spectrum that falls to zero in the last three bins of the distribution, a true spectrum in accordance with this assumption is used to generate a large number of pseudo-experiments. The pseudo-experiments undergo the fitting and unfolding procedures described in the previous sections and are used to estimate the bias of the unfolding method with respect to the true spectrum. The observed bias is used as an estimate of the type C uncertainty. As an additional check, the model dependence uncertainty is evaluated using alternative response matrices that are obtained by varying the relative fraction of the VBF and ggH components within the experimental uncertainty, as given by the CMS combined measurement~\cite{Khachatryan:2014jba}. The bias observed using this approach is found to lie within the uncertainty obtained with the method described before.

Type A and B uncertainties are finally combined together after the unfolding
summing in quadrature positive and negative contributions separately for each bin. Type C uncertainties, also referred to as ``model dependence'', are instead quoted separately.
The effect of each source of the uncertainty is quoted for each bin of \pth~in Table~\ref{table:values_and_uncertainties}.

\section{Results}
\label{sec:results}

\begin{figure}[!ht]
\centering
\includegraphics[width=0.6\textwidth]{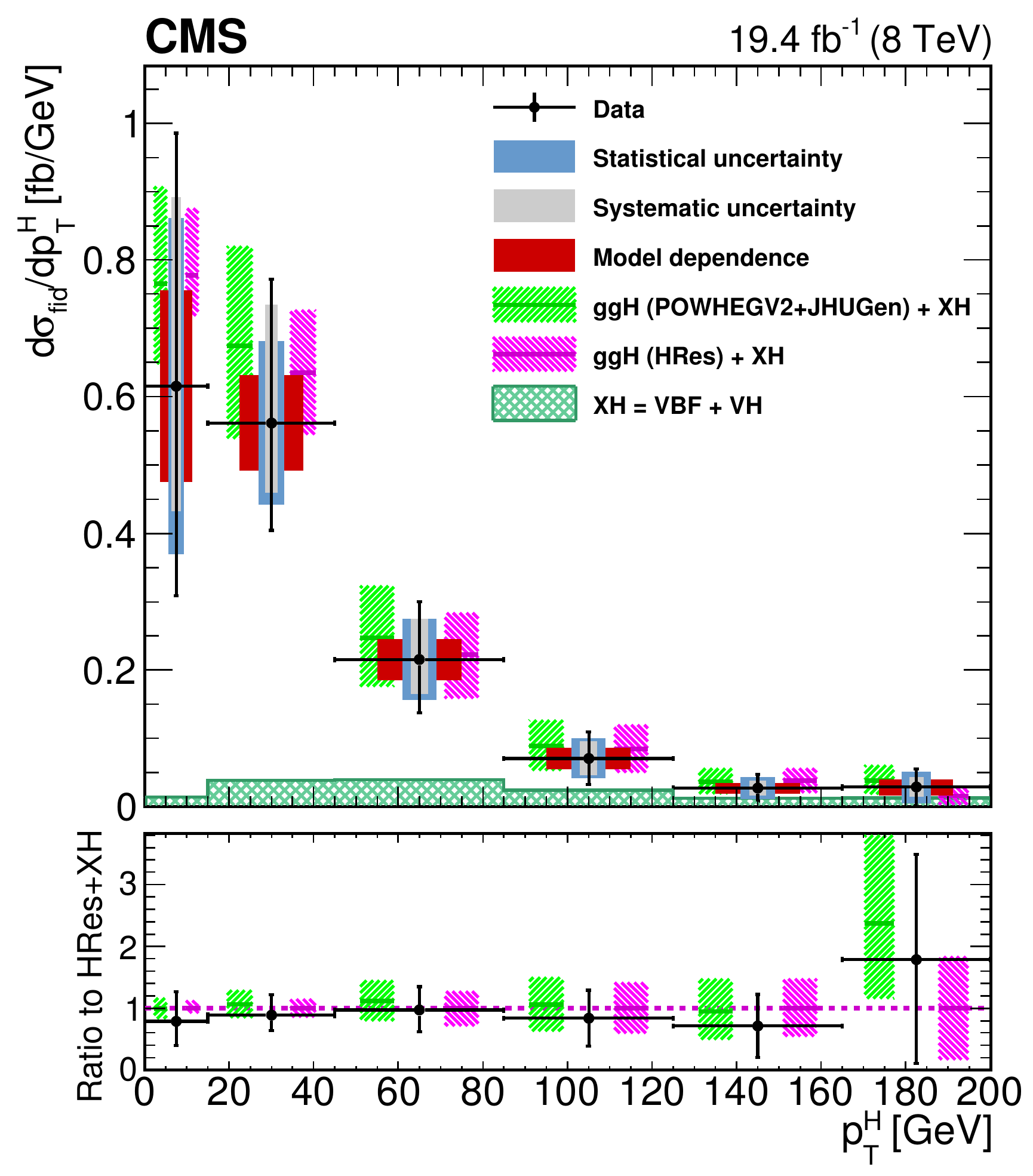}
\caption{Higgs boson production cross section as a function of \pth{}, after applying the unfolding procedure.
Data points are shown, together with statistical and systematic uncertainties. The vertical bars on the data points correspond to the sum in quadrature of the statistical and systematic uncertainties. The model dependence uncertainty is also shown.
The pink (and back-slashed filling) and green (and slashed filling) lines and areas represent the SM theoretical estimates in which the acceptance of the dominant ggH contribution is modelled by \textsc{HRes} and \POWHEG V2, respectively. The subdominant component of the signal is denoted as XH=VBF+VH and it is shown with the cross filled area separately. The bottom panel shows the ratio of data and \POWHEG V2 theoretical estimate to the \textsc{HRes} theoretical prediction.}\label{fig:unfolded}
\end{figure}

The unfolded \pth{} spectrum is shown in Fig.~\ref{fig:unfolded}. Statistical, systematic, and theoretical uncertainties are shown as separate error bands in the plot.
The unfolded spectrum is compared with the SM-based theoretical predictions where the ggH contribution is modelled using the \textsc{HRes} and \POWHEG V2 programs. The comparison shows good agreement between data and theoretical predictions within the uncertainties.
The measured values for the differential cross section in each bin of \pth{} are reported together with the total uncertainty in Table~\ref{table:values_and_uncertainties}.

\begin{table}[bh]
\renewcommand{\arraystretch}{1.1}
\topcaption{Differential cross section in each \pth{} bin, together with the total uncertainty and the separate components of the various sources of uncertainty.}\label{table:values_and_uncertainties}
\resizebox{\textwidth}{!}{
\begin{tabular}{ccccccc}
\hline
\pth&$\rd\sigma/\rd\pt^{\mathrm{H}}$&Total&Statistical&Type A&Type B&Type C\\
{[\GeVns{}]}&[fb/\GeVns{}]&uncertainty&uncertainty&uncertainty& uncertainty& uncertainty\\
&&[fb/\GeVns{}]&[fb/\GeVns{}]&[fb/\GeVns{}]&[fb/\GeVns{}]&[fb/\GeVns{}]\\
\hline
\x0--15 & 0.615 & $ {+}0.370/{-}0.307$ & $\pm$0.246 & $\pm$0.179 & $ {+}0.211/{-}0.038$  & $\pm$ 0.140\\
15--45 & 0.561 & $ {+}0.210/{-}0.157$ & $\pm$0.120 & $\pm$0.093 & $ {+}0.146/{-}0.041$  & $\pm$ 0.070\\
45--85 & 0.215 & $ {+}0.084/{-}0.078$ & $\pm$0.059 & $\pm$0.037 & $ {+}0.047/{-}0.034$  & $\pm$ 0.030\\
\x85--125 & 0.071 & $ {+}0.038/{-}0.038$ & $\pm$0.029 & $\pm$0.017 & $ {+}0.018/{-}0.017$  & $\pm$ 0.016 \\
125--165 & 0.027 & $ {+}0.020/{-}0.019$ & $\pm$0.016 & $\pm$0.009 & $ {+}0.007/{-}0.007$  & $\pm$ 0.008 \\
165--$\infty\x$ & 0.028 & $ {+}0.027/{-}0.027$ & $\pm$0.023 & $\pm$0.012 & $ {+}0.008/{-}0.007$  & $\pm$ 0.012 \\
\hline
\end{tabular}
}
\end{table}

\begin{figure}[!ht]
\centering
\includegraphics[width=0.6\textwidth]{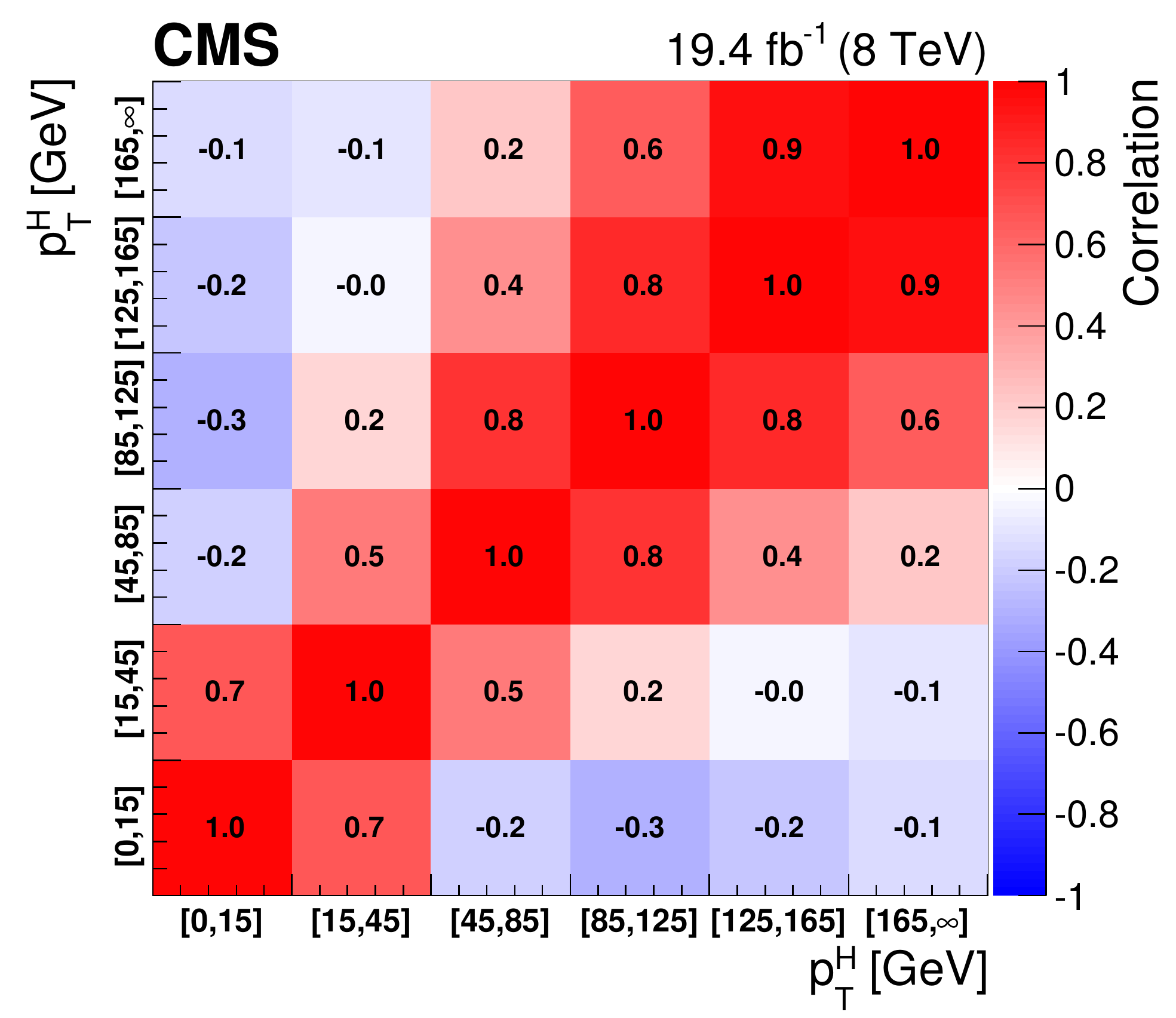}
\caption{Correlation matrix among the \pth~ bins of the differential spectrum.}\label{fig:cov_matrix}
\end{figure}

Figure \ref{fig:cov_matrix} shows  the correlation matrix for the six bins of the differential spectrum. The correlation cor($i$,$j$) of bins $i$ and $j$ is defined as:
\begin{equation}
\mathrm{cor}(i,j) = \frac{ \cov(i,j) }{  s_{i}s_{j} },
\end{equation}
where $\cov(i,j)$ is the covariance of bins $i$ and $j$, and ($s_{i}$,
$s_{j}$) are the standard deviations of bins $i$ and $j$,  respectively.

To measure the inclusive cross section in the fiducial phase space, the differential measured spectrum is integrated over \pth. In order to compute the contributions of the bin uncertainties of the differential spectrum to the inclusive uncertainty,  error propagation is performed taking into account the covariance matrix of the six signal strengths. For the extrapolation of this result to the fiducial phase space, the unfolding procedure is not needed, and the inclusive measurement has only to be corrected for the fiducial phase space selection efficiency $\epsilon_{\text{fid}}$. Dividing the measured number of events by the integrated luminosity and correcting for the overall selection efficiency, which is estimated in simulation to be $\epsilon_{\text{fid}} = 36.2 \%$, the inclusive fiducial $\sigma \, \mathcal{B}$, $\sigma_{\text{fid}}$, is computed to be:
\begin{equation}
\sigma_{\text{fid}} = 39\pm 8\stat \pm 9\syst\unit{fb},
\end{equation}
in agreement within the uncertainties with the theoretical estimate of $48 \pm 8\unit{fb}$, computed integrating the spectrum obtained with the \POWHEG V2 program for the ggH process and including the XH contribution.

\section{Summary}
\label{sec:summary}

The cross section for Higgs boson production in $\Pp\Pp$ collisions has been studied using
the ${\PH \to \PWp \PWm}$  decay mode, followed by
leptonic decays of the $\PW$ bosons to an oppositely charged
electron-muon pair in the final state.
Measurements have been performed using data from $\Pp\Pp$ collisions at a centre-of-mass energy of 8\TeV collected by the CMS experiment at the LHC and corresponding to an integrated luminosity of 19.4\fbinv.
The differential cross section has been measured as a function of the Higgs boson transverse momentum in a fiducial phase space, defined to match the experimental kinematic acceptance, and summarized in Table~\ref{table:fid_cuts}.
An unfolding procedure has been used to extrapolate the measured results to the fiducial phase space and to correct for the detector effects.
The measurements have been compared to SM theoretical estimations provided by the \textsc{HRes} and \POWHEG V2 generators, showing good agreement within the experimental uncertainties.
The inclusive production $\sigma \, \mathcal{B}$ in the fiducial phase space has been measured to be $39\pm 8\stat \pm 9\syst\unit{fb}$, consistent with the SM expectation.

\begin{acknowledgments}

We congratulate our colleagues in the CERN accelerator departments for the excellent performance of the LHC and thank the technical and administrative staffs at CERN and at other CMS institutes for their contributions to the success of the CMS effort. In addition, we gratefully acknowledge the computing centres and personnel of the Worldwide LHC Computing Grid for delivering so effectively the computing infrastructure essential to our analyses. Finally, we acknowledge the enduring support for the construction and operation of the LHC and the CMS detector provided by the following funding agencies: BMWFW and FWF (Austria); FNRS and FWO (Belgium); CNPq, CAPES, FAPERJ, and FAPESP (Brazil); MES (Bulgaria); CERN; CAS, MoST, and NSFC (China); COLCIENCIAS (Colombia); MSES and CSF (Croatia); RPF (Cyprus); SENESCYT (Ecuador); MoER, ERC IUT and ERDF (Estonia); Academy of Finland, MEC, and HIP (Finland); CEA and CNRS/IN2P3 (France); BMBF, DFG, and HGF (Germany); GSRT (Greece); OTKA and NIH (Hungary); DAE and DST (India); IPM (Iran); SFI (Ireland); INFN (Italy); MSIP and NRF (Republic of Korea); LAS (Lithuania); MOE and UM (Malaysia); BUAP, CINVESTAV, CONACYT, LNS, SEP, and UASLP-FAI (Mexico); MBIE (New Zealand); PAEC (Pakistan); MSHE and NSC (Poland); FCT (Portugal); JINR (Dubna); MON, RosAtom, RAS and RFBR (Russia); MESTD (Serbia); SEIDI and CPAN (Spain); Swiss Funding Agencies (Switzerland); MST (Taipei); ThEPCenter, IPST, STAR and NSTDA (Thailand); TUBITAK and TAEK (Turkey); NASU and SFFR (Ukraine); STFC (United Kingdom); DOE and NSF (USA).

Individuals have received support from the Marie-Curie programme and the European Research Council and EPLANET (European Union); the Leventis Foundation; the A. P. Sloan Foundation; the Alexander von Humboldt Foundation; the Belgian Federal Science Policy Office; the Fonds pour la Formation \`a la Recherche dans l'Industrie et dans l'Agriculture (FRIA-Belgium); the Agentschap voor Innovatie door Wetenschap en Technologie (IWT-Belgium); the Ministry of Education, Youth and Sports (MEYS) of the Czech Republic; the Council of Science and Industrial Research, India; the HOMING PLUS programme of the Foundation for Polish Science, cofinanced from European Union, Regional Development Fund; the Mobility Plus programme of the Ministry of Science and Higher Education (Poland); the OPUS programme, contract Sonata-bis DEC-2012/07/E/ST2/01406 of the National Science Center (Poland); the Thalis and Aristeia programmes cofinanced by EU-ESF and the Greek NSRF; the National Priorities Research Program by Qatar National Research Fund; the Programa Clar\'in-COFUND del Principado de Asturias; the Rachadapisek Sompot Fund for Postdoctoral Fellowship, Chulalongkorn University (Thailand); the Chulalongkorn Academic into Its 2nd Century Project Advancement Project (Thailand); and the Welch Foundation, contract C-1845.

\end{acknowledgments}

\clearpage

\bibliography{auto_generated}

\cleardoublepage \appendix\section{The CMS Collaboration \label{app:collab}}\begin{sloppypar}\hyphenpenalty=5000\widowpenalty=500\clubpenalty=5000\input{HIG-15-010-authorlist.tex}\end{sloppypar}
\end{document}

%% file: HIG-15-010-authorlist.tex
\textbf{Yerevan Physics Institute,  Yerevan,  Armenia}\\*[0pt]
V.~Khachatryan, A.M.~Sirunyan, A.~Tumasyan
\vskip\cmsinstskip
\textbf{Institut f\"{u}r Hochenergiephysik der OeAW,  Wien,  Austria}\\*[0pt]
W.~Adam, E.~Asilar, T.~Bergauer, J.~Brandstetter, E.~Brondolin, M.~Dragicevic, J.~Er\"{o}, M.~Flechl, M.~Friedl, R.~Fr\"{u}hwirth\cmsAuthorMark{1}, V.M.~Ghete, C.~Hartl, N.~H\"{o}rmann, J.~Hrubec, M.~Jeitler\cmsAuthorMark{1}, A.~K\"{o}nig, I.~Kr\"{a}tschmer, D.~Liko, T.~Matsushita, I.~Mikulec, D.~Rabady, N.~Rad, B.~Rahbaran, H.~Rohringer, J.~Schieck\cmsAuthorMark{1}, J.~Strauss, W.~Treberer-Treberspurg, W.~Waltenberger, C.-E.~Wulz\cmsAuthorMark{1}
% ...
\vskip\cmsinstskip
% ...
1:~~Also at Vienna University of Technology, Vienna, Austria\\
% ...
